\documentclass[%
reprint,
showpacs,preprintnumbers,
 amsmath,amssymb,
 aps,pre,
]{revtex4-1}

\usepackage{graphicx}
\usepackage{dcolumn}
\usepackage{bm}

\newcommand{\Dmax}{D_\textrm{max}}
\newcommand{\Tmax}{T_\textrm{max}}
\newcommand{\Flat}{F_\textrm{lat}}
\newcommand{\overbar}[1]{\mkern 0.5mu\overline{\mkern-0.5mu#1\mkern-1.5mu}\mkern 1.5mu}

\begin{document}


\title{Temperature-abnormal diffusivity in underdamped
space-periodic systems driven by external time-periodic force}

\author{I.G. Marchenko}
\affiliation{%
 NSC \lq\lq Kharkov Institute of Physics and Technology\rq\rq,
 1 Akademicheskaya str., Kharkov 61108, Ukraine
}%
\affiliation{
Kharkov National University,
4 Svobody Sq., Kharkov 61077, Ukraine%
}%
\author{I.I. Marchenko}
\affiliation{
 NTU \lq\lq Kharkov Polytechnic Institute\rq\rq,
 21 Frunze str., Kharkov 61145, Ukraine
}%
\author{A.V. Zhiglo}\email{azhiglo@uchicago.edu}
\affiliation{%
 NSC \lq\lq Kharkov Institute of Physics and Technology\rq\rq,
 1 Akademicheskaya str., Kharkov 61108, Ukraine%
}%
\affiliation{
Kavli Institute for the Physics and Mathematics of the Universe (WPI), 
The University of Tokyo, Kashiwa, Chiba 277-8583, Japan%
}%
\date{\today}

\begin{abstract}
We present a study of diffusion enhancement of underdamped Brownian
particles in 1D symmetric space-periodic potential due to external
symmetric time-periodic forcing with zero mean. We show that the diffusivity
can be enhanced by many orders
of magnitude at appropriate choice of the forcing amplitude and frequency.
The diffusivity demonstrates TAD, abnormal (decreasing) temperature
dependence at forcing amplitudes exceeding certain value.
At any fixed forcing frequency $\Omega$ normal temperature dependence of the
diffusivity is restored at low enough temperatures, $T<T_\textrm{TAD}(\Omega)$
--- in contrast with the problem with constant external forcing.
At fixed temperature at small forcing frequency the diffusivity
either slowly decreases with $\Omega$, or (at stronger forcing)
goes through a maximum near $\Omega_2$, reciprocal superdiffusion
stage duration. At high frequencies, between $\Omega_2$ and a fraction of
the oscillation frequency at the potential minimum, the diffusivity is shown to
decrease with $\Omega$ according to a power law, with exponent related to the
transient superdiffusion exponent. This behavior is found similar for the cases
of sinusoidal in time and piecewise constant periodic (``square'') forcing.
\end{abstract}

\maketitle


\section{\label{s:intro}INTRODUCTION}

The phenomena of diffusion and transport over a potential energy landscape
play a key role in a number of
processes in physics, chemistry and biology
\cite{MehrerDifsnSolids,AntczakSurfaceDiffusion,*Enzymes,
*SChakrabortyMicrofluid,
BraunKivsharFrenkelKont,*Colloid1Dsin,*GoychukNanoMolec,
*CellTransport,Hanggi09RMP,Evers13ColloidsLight}.
Josephson tunneling junctions, superionic conductors, phase-locked-loop
frequency control systems, charge density waves are a few examples of
systems in which these processes in periodic potential are important
\cite{Riskenbook}.

In recent years the interest has been growing to experimental studies of
manipulating the particle diffusion by external fields. One can effectively
control the diffusion processes by varying the field parameters. For instance,
huge enhancement in diffusivity (exceeding its value in the same system
without extra forcing by a factor of order hundreds) was observed in studying
particle diffusion in colloids with optical traps (optical vortices)
\cite{Lee06PRL}. Large increase in diffusivity was also observed in studies
of paramagnetic particle diffusion on surfaces of garnet ferrites influenced
by external time-periodic magnetic field \cite{Tierno10PRL}.
In the same manner diffusivity growth with shaking strength was noticed in
experiments with granular gas \cite{Eshuis10PRL}. It turned possible to
increase the diffusivity of ions in membrane channels by varying the external
electromagnetic field \cite{Pagliara13AdvMat}. One can increase the rates of
diffusion-limited physical processes, and effectively separate micro- and
nano-particles of different nature by varying diffusion coefficients in
different directions \cite{Sajeesh13MicNano}. 

Despite these remarkable
achievements, quantitative description of such enhanced diffusion under
the influence of external forces remains fragmentary to date.

\subsection{Diffusion of Brownian particles in spatially-periodic potentials
under constant external forcing}
Special cases of underdamped and overdamped physical systems are often
considered. In overdamped systems inertial effects may be neglected,
significantly
simplifying the mathematical treatment of the problem. In under\-damp\-ed
systems on the other hand the viscous decay of oscillations occurs on times
long compared with the oscillation period in the potential; such systems are
harder to investigate, and the progress in studying these is more limited
compared with overdamped ones.

First detailed studies of a Brownian particle motion in a washboard potential
(\emph{i.e.} a periodic potential tilted with an additional
constant force field) were carried out by H.~Risken and colleagues
\cite{Vollmer79ZPhB,*Risken79ZPhB,
79RiskenPhLtA,*Risken79ZPhBnlin,*Vollmer80ZPhB,*Jung84ZPhB}, for all values
of the friction.
It was shown that at low friction the appearance of
\lq\lq locked\rq\rq\ (in which the particle oscillates around one minimum of
the potential) and \lq\lq running\rq\rq\ (where the particle travels
through many potential periods, not getting trapped at separate minima)
solutions was important for the particle ensemble behavior. At small
temperatures running solutions appear abruptly at the external force value
exceeding certain critical $F_\textrm{cr}$;
at smaller forces a single class of solutions (locked) is realized.
Systematization of the results of these studies may be found in authoritative
monograph \cite{Riskenbook}.

Approximate solution of the Fokker-Planck equation (FPE) for the coordinate and 
velocity distribution function was the central moment of these studies.
Two different approaches were used, that were checked to produce compatible 
results. 

In the first of these 
\cite{Vollmer79ZPhB,*Risken79ZPhB} by expanding the distribution function in
Hermite functions in velocity and Fourier-series in space coordinate,
the Fokker-Planck equation was transformed into an infinite set of coupled
equations for the expansion coefficients.
These equations were solved by a matrix continued fraction method.
Particle flux was then expressed through the expansion coefficients, after
appropriate series truncation. At smaller friction progressively more 
terms must be kept, making this method less efficient.

Another approach 
\cite{79RiskenPhLtA,*Risken79ZPhBnlin,*Vollmer80ZPhB,*Jung84ZPhB} is best 
suited for low friction, when the energy becomes a slow variable (changing
on friction dissipation timescales). The distribution function was rewritten
in the energy and configuration space coordinate variables, FPE
was solved in these new variables with the help of physically-motivated 
self-consistent ansatz.

Many illuminating results were obtained for the distribution functions at 
various parameter values, stability domains of locked and running solutions, 
plots of the mean particle velocity, etc. Diffusion properties however 
remained not studied. This matter was partly addressed in other works, 
based on simulations of the stochastic Langevin equation.

Early systematic study of diffusion by means of numerical simulation of
the Langevin equation in periodic potential with extra constant forcing was
undertaken in \cite{Costantini99EPL} (similar simulations were reported in
\cite{Gang96PRL,*Borromeo99PRL} for forcing with a time-periodic component).
Costantini \& Marchesoni analyzed both overdamped and
underdamped settings. At low dissipation they observed
significant enhancement of the spatial diffusion near the critical force
$F_\textrm{cr}$ (\cite{Costantini99EPL}, Fig.~1: diffusivity
ten times larger than its post-maximum
value at $F=2F_\textrm{cr}$, at the temperature equal to 0.6 of the
potential well depth $U_0$), and related that to features of the particle jump
statistics and locked-to-running transition.

Further progress in studies of the diffusion under the action of constant force
was made in \cite{Lindenberg07PRL,Khoury09PRE,*Sancho04PRL,Lindenberg05NJPh}.
Time-dependence of the particle ensemble dispersion was studied in
\cite{Lindenberg07PRL}. It was shown for the first time that in underdamped
systems a special regime of dispersionless transport was realized, in which
dispersion virtually does not change with time, on certain limited interval
of time. The authors explained this phenomenon, they showed that strongly
non-equilibrium distribution of particles in space (with steep front and
exponential tail; formed as the particles first exit their original potential
well) persists for long times, before eventually broadening and assuming normal
Gaussian shape.

The temperature dependence of the diffusivity was studied as well
\cite{Lindenberg05NJPh}. The authors observed that the maximal diffusivity
$D_\textrm{max}$ was achieved near the $F_\textrm{cr}$.
At low friction $D_\textrm{max}$ was shown to depend abnormally on the
temperature --- increase with decreasing temperature $T$. Fitting
(for the three temperatures studied) yielded
$D_\textrm{max}\propto T^{-3.5}$  relation.

The presence of such abnormal temperature dependence is one important
aspect in which underdamped systems differ from over\-damp\-ed ones. In fact,
certain peculiarities in diffusivity at low temperatures are observed (for
driven Brownian particles in space-periodic potentials) in overdamped situation
as well. P. Reimann and colleagues
\cite{Reimann01PRL,*Reimann02PRE} showed analytically that for a sinusoidal in
space potential at external constant force value near $F_0$, the value
required for direct particle pull over the potential barrier
(Eq.~\ref{Eq:Flat} below, p.~\pageref{Eq:Flat}), the ratio of $\Dmax$ to the
diffusivity
value $D_0$ in the viscous medium without lattice and bias forces grows at
temperature decreasing $\propto T^{-2/3}$. Yet the diffusivity itself still
vanishes at $T\to 0$, as $\Dmax\propto T^{1/3}$ \cite{Reimann02PRE}, and grows
monotonically for all $T$.

Regions with $\partial D/\partial T<0$ were obtained in overdamped problem in
\cite{Lindner01FlucNoise} (cf. Fig.~6) for specially crafted exotic potential,
flat apart from narrow peaks. The authors linked this abnormal diffusivity
behavior to the large ratio of relaxation to escape time in such a system.
Similar abnormal $D(T)$ behavior was found in \cite{Heinsalu04}
(cf. Fig.~2--4) for piecewise linear potential at extreme asymmetry,
essentially sawtooth. So, while possible at large friction,
special atypical conditions must be met for existence of parameters $\{F,T\}$
at which
$\partial D/\partial T<0$. On the contrary, such parameters exist universally
in the problem with small friction (\cite{MarchenkoFconst}, more below).

Another distinction of the diffusivity in overdamped problem is that the
maximum in $D(F)$ is achieved close to the $F_0$. Whereas at low friction
$\Dmax(F)$ is achieved at (typically much smaller) $F_\mathrm{cr}$.
Experimental data for diffusion of colloidal particles in a tilted potential
created with laser traps \cite{Evstigneev08PRE}, Fig.~1, agreed reasonably
well with this analytic \cite{Reimann02PRE} $D(F)$ dependence. The potential
barrier height and the free diffusion coefficient were treated as fit
parameters.
Similarly, diffusivity in \cite{Ma15ColloidDifsn} (Fig.~7) showed peaked at
$F_0$ behavior, and agreed fairly well with analytic predictions from
\cite{Reimann02PRE} with the potential form inferred from independent study
of the colloidal particle spatial distribution function. In this work colloidal
particles diffused over the potential created by the bottom grid of colloidal
spheres, and extra constant forcing was due to gravity (controlled by the
sample inclination angle). Different limiting cases of analytic predictions
\cite{Reimann02PRE} for the mean particle velocity and diffusivity were
compared with experimental results at different settings, and showed good
agreement.

The physical reason behind the abnormal temperature dependence of diffusivity
in underdamped systems is traced to increasing jump length of running particles
(before getting retrapped at another potential minimum) with the temperature
decrease \cite{MarchenkoFconst}. This phenomenon being absent in overdamped
systems (as these have no proper running states).
The maximal diffusivity is achieved at forcing near $F_\mathrm{cr}$, as at
this value populations of locked and running particles are about equal, and it
is the mutual motion between these two populations (with
temperature-independent running particle speed) that leads to fast spreading
of the particle packet, manifested as giant diffusion \cite{Lindenberg05NJPh}.

More thorough analysis of (the temperature- and force-dependence of) the
diffusion in underdamped systems was carried out in \cite{Marchenko12EPL}.
It was shown by numeric simulations of the Langevin equation
that the phenomenon of diffusivity growth with the temperature
decreasing (in the current work called TAD, ``temperature-abnormal
diffusivity'') was only manifested in a narrow interval of applied external
forcing. Study in a wide range
of temperatures demonstrated that the diffusivity grew with decreasing
temperature
as $\Dmax\propto T^{2/3}\exp{[\mathcal{E}/(k_BT)]}$ for certain $\mathcal{E}>0$,
checked valid for
$k_B T/U_0\in[0.05;0.85]$; whereas the power-law fit
$D_\textrm{max}\propto T^{-3.5}$ proposed in \cite{Lindenberg05NJPh} could
approximate the results only in the narrow temperature range studied by the
authors.
We showed that the main exponential growth resulted from such of the correlation
time at the temperature decreasing (found from the simulation results).

The diffusion properties were further elucidated and systematized for a broad
range of the forces $F$, friction coefficients $\gamma$ and temperatures $T$
in \cite{MarchenkoFconst}, in sinusoidal 1D
potential in underdamped setting. We introduced
a velocity potential $W(V)$, which in part governed time evolution of certain
slow (not changing much on times of order one period
$\widetilde{\mathcal{T}}_0$ of small oscillations in the
potential, rather evolving on viscous dissipation timescales) velocity
variable, defined through a discrete quantity, maximal velocity in such
consecutive oscillations. With
certain reservations that $W(V)$ also defined the velocity distribution
function $N(V)$ agreeing well with the one found numerically. The latter
showed clear bimodal structure with maxima at $V=0$ (locked states) and
$V=F/\gamma$ (running ones) at intermediate forcing values. In this region
of forcing values TAD was observed. 

Analytic expressions for the mean particle velocity
$\langle V\rangle$ and diffusivity were obtained with the model $N(V)$, good
agreement was shown with direct Langevin equation simulation results.
Simple fits were proposed determining the model
parameters [entering $W(V)$ and $N(V)$] through $F$ and $\gamma$, which
--- apart from usefulness for experimentalists/applications --- allowed
one to comprehend the features of transport and diffusivity in all the
physical parameter space. Physical reason for the
$\Dmax\propto\exp{[\mathcal{E}/(k_BT)]}$ scaling was explained. More detail
in the introductory part of Sec.~\ref{s:results},
p.~\pageref{s:results} below.

\subsection{Periodic in time external forcing}
In practical applications periodic in time external forces (electromagnetic,
acoustic, etc.) are more straightforwardly realized
than constant forcing. 
It is natural to study how the findings for the constant forcing problem are
modified for the time-periodic forcing setup, both from theoretical and
practical point of view. By changing the diffusivity with the aid of periodic
external fields it is possible to create surface structures with prescribed
properties; influence the dynamics of point defects, under irradiation in part;
alter the dislocation motion dynamics; increase chemical reaction rates in
selected directions; manipulate the efficiency of processes in
biological systems, to mention just a few applications
\cite{DynamicsSolidSurface,*ChemPhysicsSurface,Kulemin
}.

Most of the works on periodic forcing deal with ratchet-type systems
\cite{Hanggi09RMP} or stochastic resonance. Studies focusing on diffusion
properties are scarce. We are not aware of any works that systematically
study the combined dependence of the diffusivity on the temperature, the
forcing amplitude and frequency in a representatively broad range of these
parameters.

Dependence of the diffusivity on the interplay between the
external periodic (piecewise constant, Eq.~\ref{F:PCP} on p.~\pageref{F:PCP}
below) forcing and the lattice
force was studied in \cite{Speer12EPL} in 2D overdamped setup. The forcing
frequency was fixed at one value.
It was shown that in a system with a sum of Yukawa potentials at square
grid nodes the particle diffusivity was significantly
enhanced for certain intricate set of bands (``A-windows'') of the forcing
amplitudes and orientations with respect to the grid. At special
forcing arrangement TAD was observed for some diffusivity tensor components,
and one of its principal values increased exponentially with inverse
temperature. The other diffusivity tensor principal value decreased
exponentially at the same time. It is likely that multi-dimensionality effects
were crucial for TAD. Instead of having two, locked and running, populations in
1D accounting for (already sophisticated in this case) diffusion features,
flows in higher dimensional problems are far more complex, known to result in
further peculiarities in transport and diffusion
\cite{Evers13ColloidsLight,Sancho10EPJST}.

Underdamped 1D system was studied in \cite{Marchenko12JETPL}, with a sinusoidal
potential (same as Eq.~\ref{Ux} below) and sinusoidal in time forcing.
Diffusivity enhancement by orders of
magnitude was seen. TAD was observed for some (low) forcing frequencies, at
the forcing amplitude $0.15F_0$ studied. Dependence of the diffusivity on the
forcing frequency was studied in more detail in \cite{Marchenko14JPhCS}, and
was shown to be nonmonotonic at large forcing amplitudes ($0.15F_0$ and
$0.25F_0$, at $T=240$~K, $k_B T/U_0=0.258$.) At $0.15F_0$ the maximum in the frequency dependence
of the
diffusivity $D(\Omega)|_{\{T,F\}=const}$ persisted for all the temperatures studied, corresponding frequency
$\Omega_\mathrm{max}$ increasing from $\sim 2\times 10^{-4}$ to
$\sim 6\times 10^{-3}$ (in units of the potential eigenfrequency) at $T$
varying from 120 to 360~K, $k_B T/U_0\in[0.129; 0.388]$.


The goal of the present work is to investigate in detail the dependence of the
diffusivity on the amplitude and the frequency of uniform time-periodic
external forcing in a 1D underdamped space-periodic system, and conditions
under which TAD is realized. We
mainly study piecewise constant periodic (PCP: Eq.~\ref{F:PCP}) forcing as this
case is more straightforward to relate to the results for constant forcing.

The equations solved and the methods used are described in Sec.~\ref{s:setup}.
In Sec.~\ref{s:results} we show how the diffusivity changes with the PCP
forcing frequency $\Omega$, starting from the understood results at constant
forcing, $\Omega=0$. We observe that TAD is realized for some
(intermediate) forcing amplitudes. Sec.~\ref{s:DvsT} is devoted to
studying the $D(T)$ behavior at fixed $\Omega$. The results for the diffusivity
in the case of sinusoidal in time forcing are shown in Sec.~\ref{s:Fsin}.
We conclude in Sec.~\ref{s:concl}.
\section{PROBLEM SETUP AND NUMERICAL METHOD}\label{s:setup}
\subsection{Problem setup}\label{ss:setup}
In this paper the 1D dynamics of a particle subject to external forcing $F_t$
is described by the Langevin equation:
\begin{equation}\label{Langevin}
  m\ddot{x}=-
  dU(x)/dx-\gamma\dot{x}+F_t(t)+\xi(t),
\end{equation}
where $t$ is the time, $x$ is the particle coordinate, $m$ is its mass,
$\gamma$  is the friction coefficient. Overdot stands for time
differentiation. $\xi(t)$ represents thermal fluctuations, described
by Gaussian white noise with correlation
\begin{equation}\label{xi_cor}
  \langle\xi(t)\xi(t^\prime)\rangle=2\gamma k_B T \delta(t-t^\prime);
  \; \langle\xi(t)\rangle=0.
\end{equation}
Here  $k_B$ is the Boltzmann constant, $T$  is the temperature.

The potential energy $U(x)$ is taken
\begin{equation}\label{Ux}
  U(x)=-(U_0/2) \cos(2\pi x/a),
\end{equation}
where $a$ is the lattice constant and  $U_0$ is the potential barrier height.

The lattice force  $\Flat$ acting upon the particle is
\begin{equation}\label{Eq:Flat}
  \Flat (x)=-dU/dx = -F_0\sin(2\pi x/a).
\end{equation}
Quantity $F_0=\pi U_0/a$  corresponds to the minimal external force needed
to drag the particle over the potential barrier separating potential minima
on the 1D lattice at large friction.

We use the same physical parameters as in
\cite{
MarchenkoFconst}, typical for adatom diffusion on close-packed
metal surfaces. Namely, values
\begin{equation}\label{U0a}
  U_0=80\mathrm{\ meV,}\; a = 2 \mathrm{\ \AA}
\end{equation}
for the activation barrier and the lattice constant are adopted.

Two possibilities for the external periodic forcing are studied: piecewise
constant periodic ({\bf PCP}) driving
\begin{equation}\label{F:PCP}
  F_t(t)=F_e\,\mathrm{sign}\,\![\sin(\Omega t)]
\end{equation}
and sinusoidal one,
\begin{equation}\label{F:sin}
  F_t(t)=F_e \sin (\Omega t) .
\end{equation}
$\Omega$ is the angular frequency of the external force and  $F_e$ is its
amplitude.

It is convenient to use dimensionless time $t^\prime$ and space coordinate
$x^\prime$, normalizing the physical values to the period of small oscillations
$\mathcal{T}_0\equiv2\pi/\Omega_0=a(2m/U_0)^{1/2}$ at the potential minimum
in the absence of friction,
and to the lattice spatial period $a$ respectively:
\begin{equation}\label{dimtx}
  t^\prime=\frac t{\mathcal{T}_0}=\frac t a \sqrt\frac{U_0}{2m} ,
  \quad x^\prime=\frac x a .
\end{equation}

Dimensionless temperature $T^\prime$ and friction coefficient
$\gamma^\prime$ are further introduced:
\begin{equation}\label{dimTgamma}
  T^\prime=2k_B T/U_0,\; \gamma^\prime=\gamma a(mU_0/2)^{-1/2}.
\end{equation}
Definitions (\ref{dimtx},\ref{dimTgamma}) differ from such in
\cite{Lacasta04sdif,Lindenberg07PRL,Marchenko12EPL,MarchenkoFconst,%
Marchenko12JETPL,Marchenko14JPhCS}
by extra factors of 2 appearing in the definition of $t^\prime$, and
consequently in $T^\prime$ and $\gamma^\prime$ as well. Current definition
makes interpreting the results easier ($t^\prime=1$ corresponds to one
oscillation period at small $\gamma^\prime$).
For more straightforward comparison with our previous
works we also use another dimensionless temperature in the figures,
\[ T^\dagger=k_B T/U_0=T^\prime/2.\]
The parameters of the dimensionless problem are $F_e/F_0$,
$\omega=\Omega\mathcal{T}_0$, $\gamma^\prime$, $T^\prime$. In the new
variables the equation of motion has form Eq.~1 with unit new mass, 
potential $\cos(2\pi x^\prime)$ with no prefactors and the external forcing
amplitude made dimensionless with a $2a/U_0$ multiplier.

Below lowercase $\omega$'s (frequencies), and corresponding periods
$\tau=2\pi/\omega$ denote dimensionless quantities. Physical frequencies and
periods are obtained by respectively dividing and multiplying these by the
the period $\mathcal{T}_0$ of small oscillations in the lattice potential
at $\gamma=0$.

Overdamped problem is characterized by $\gamma/m\gg \Omega_0$, whereas
$\gamma/m\ll \Omega_0$ (equivalently, $\gamma^\prime\ll 1$) corresponds
to underdamped dynamics.
We study the underdamped case here. Friction coefficient $\gamma^\prime=0.2$
we use corresponds to the same physical value as in earlier studies
\cite{Marchenko12EPL,MarchenkoFconst}.

\subsection{Numerical method}%
We solve stochastic equation (\ref{Langevin},\ref{xi_cor}) numerically
using a Verlet-type algorithm \cite{Kuznetsov} with a time step $\Delta t$
somewhat shorter than $1/100^\textrm{th}$ of the period of oscillation
$\mathcal{T}_0$. The statistic averaging is performed over the ensemble
consisting of at least $N=5\times 10^4$ particles. To verify the modeling
consistency some computations were performed with  $N=5\times 10^6$.

The initial conditions are set as follows. Each particle is placed at $x = 0$,
its velocity chosen at random, with Maxwellian distribution for a given $T$.
Thermalization over 100 oscillation periods, in the potential $U(x)$ but
without the $F_t(t)$, is then used to get equilibrium particle distribution
over both coordinate and velocity. Tests show that the distribution
function does not change after that time. In the process of such thermalization
the particles can jump to nearby elementary cells of the lattice.
Such relocated particles are returned back to the initial cell by translation
over an integer multiple of the lattice constant, to get diffusion
of all particles starting from the initial cell of the lattice.

The diffusivity is computed based on the dispersion $\sigma^2$ of the particle
distribution in the limit of large times:
\begin{equation}\label{dsigma}
  D^\prime=\lim\limits_{t\to\infty}D_\textrm{eff}^\prime(t)=
  \lim\limits_{t^\prime\to\infty}
  \frac{\bm{\langle}(x^\prime-\langle x^\prime\rangle)^2\bm{\rangle}
  (t^\prime)}%
  {2t^\prime}  \equiv\lim\limits_{t^\prime\to\infty}\frac{\sigma^2}{2t^\prime}.
\end{equation}
Angle brackets $\langle\ldots\rangle$ with no subscripts mean averaging over
the ensemble.
$D^\prime$ and $\sigma$ are dimensionless in our notation. The physical
diffusivity is $D=D^\prime a^2/\mathcal{T}_0$.

For each diffusivity calculation we find time $t^\prime_\mathrm{lin}$, after
which the dispersion grows linearly with time (if averaged over the forcing
period). The $D^\prime$ is calculated as $\sigma^2/(2t^\prime)$ at
$t^\prime=100t^\prime_\mathrm{lin}$.

It is known that in systems with low dissipation transient regimes of anomalous
diffusion can be realized \cite{Lacasta04sdif}, characterized
by
\begin{equation}\label{sdifalpha}
  \sigma^2=\bm{\langle}(x^\prime-\langle x^\prime\rangle)^2\bm{\rangle}
  \propto t^{\prime\alpha}.
\end{equation}
The process with $\alpha >1$ is referred to as superdiffusion, whereas
subdiffusion is characterized by $\alpha <1$.
Stages of anomalous diffusion were observed in our simulations. The choice
of initial conditions above aided in having these stages in clear form.

We determine the exponents $\alpha$ by the least-square linear fitting between
$\log(\sigma^2)$ and  $\log t$ on the time-interval of interest.
\section{Piecewise constant periodic forcing. Dependence of the diffusivity
on the forcing frequency}\label{s:results}
In \cite{Marchenko12EPL} it was shown that
depending on the value of the external force (constant in that study)
different functional dependences of the diffusivity on the temperature were
realized.

\begin{figure}[htbp!]
  \begin{center}
    \includegraphics[width=0.47\textwidth]{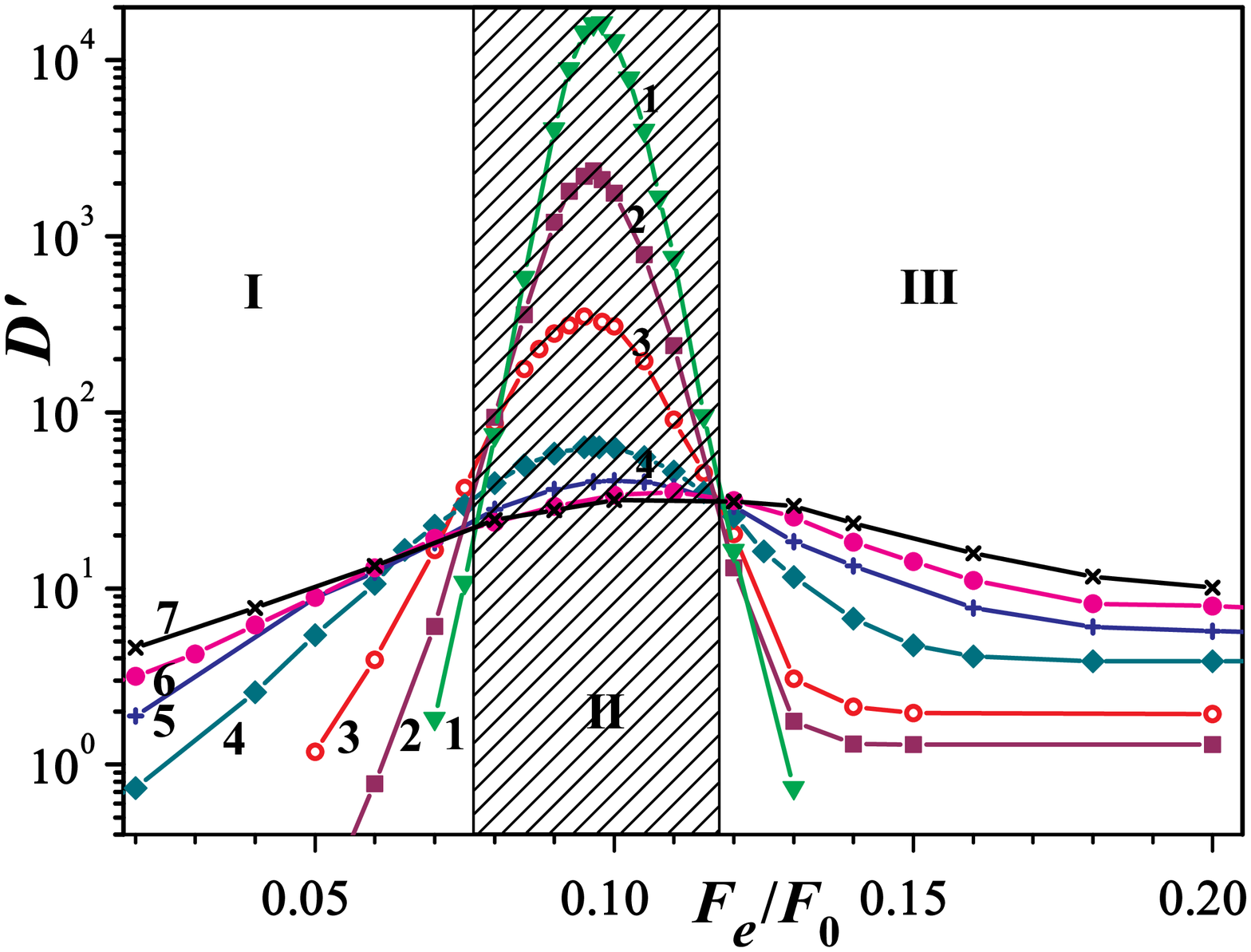}\\
     \includegraphics[width=0.123\textwidth]{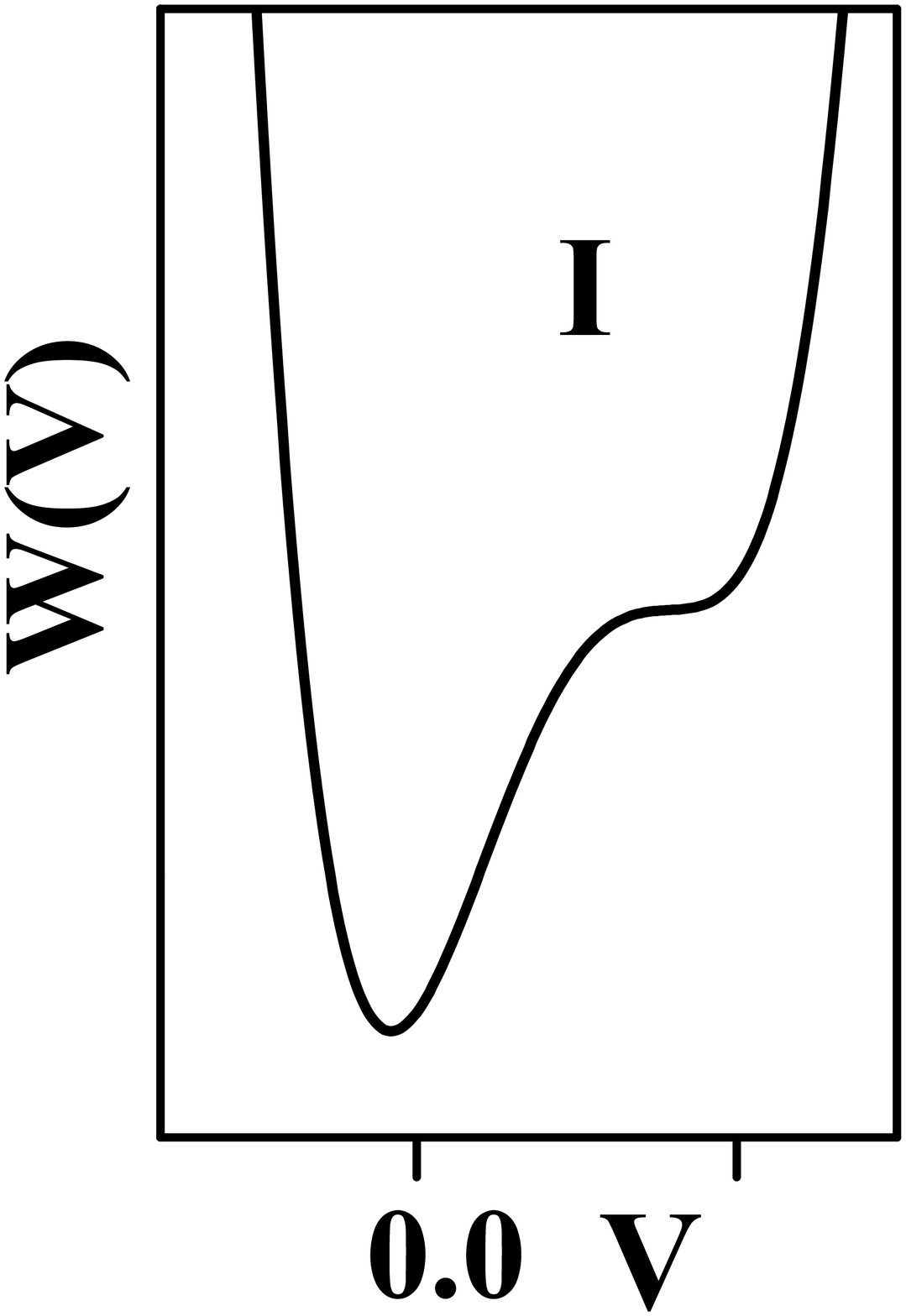} $\qquad\:$
     \includegraphics[width=0.1\textwidth]{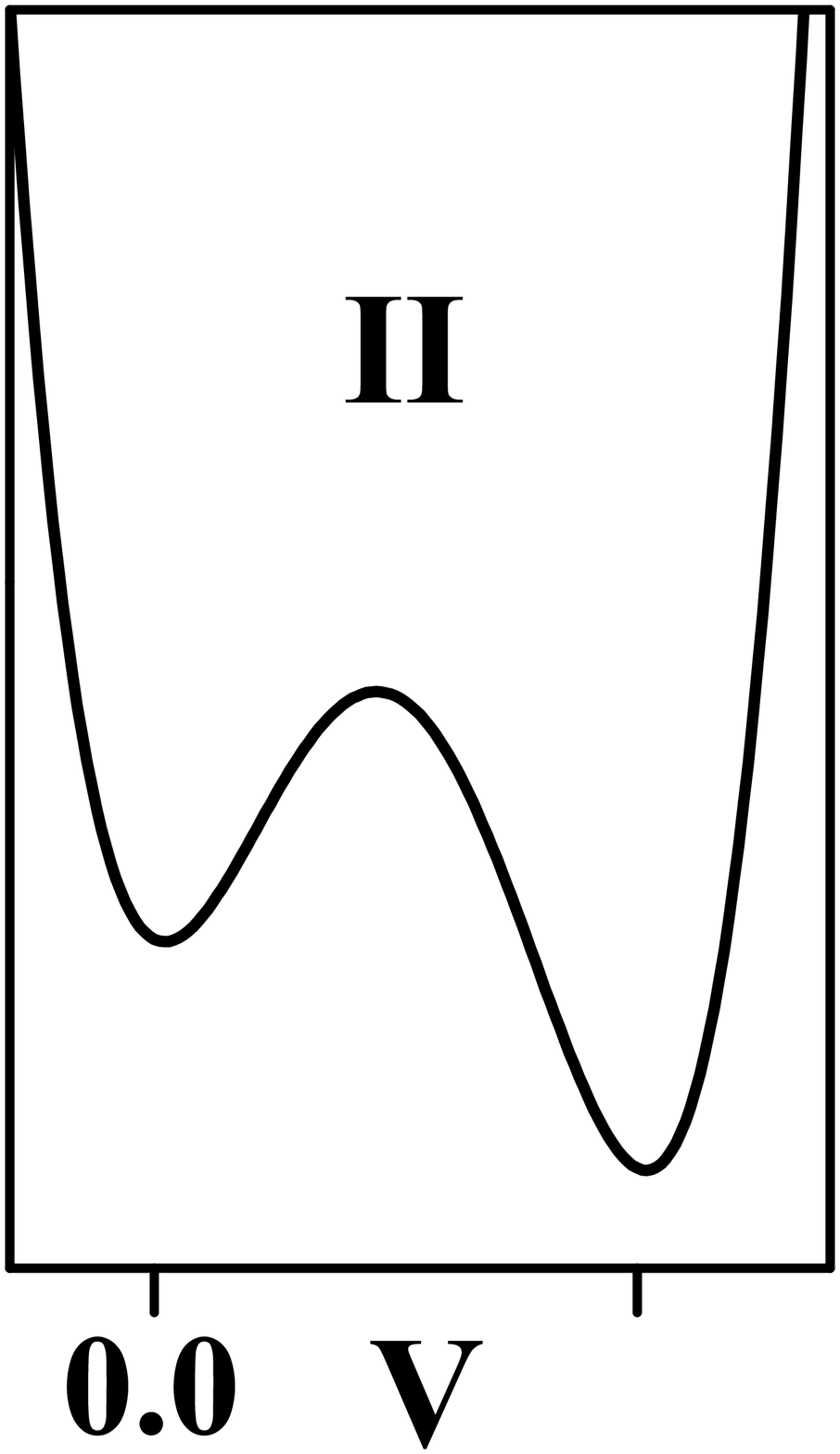} $\qquad\:$
     \includegraphics[width=0.1\textwidth]{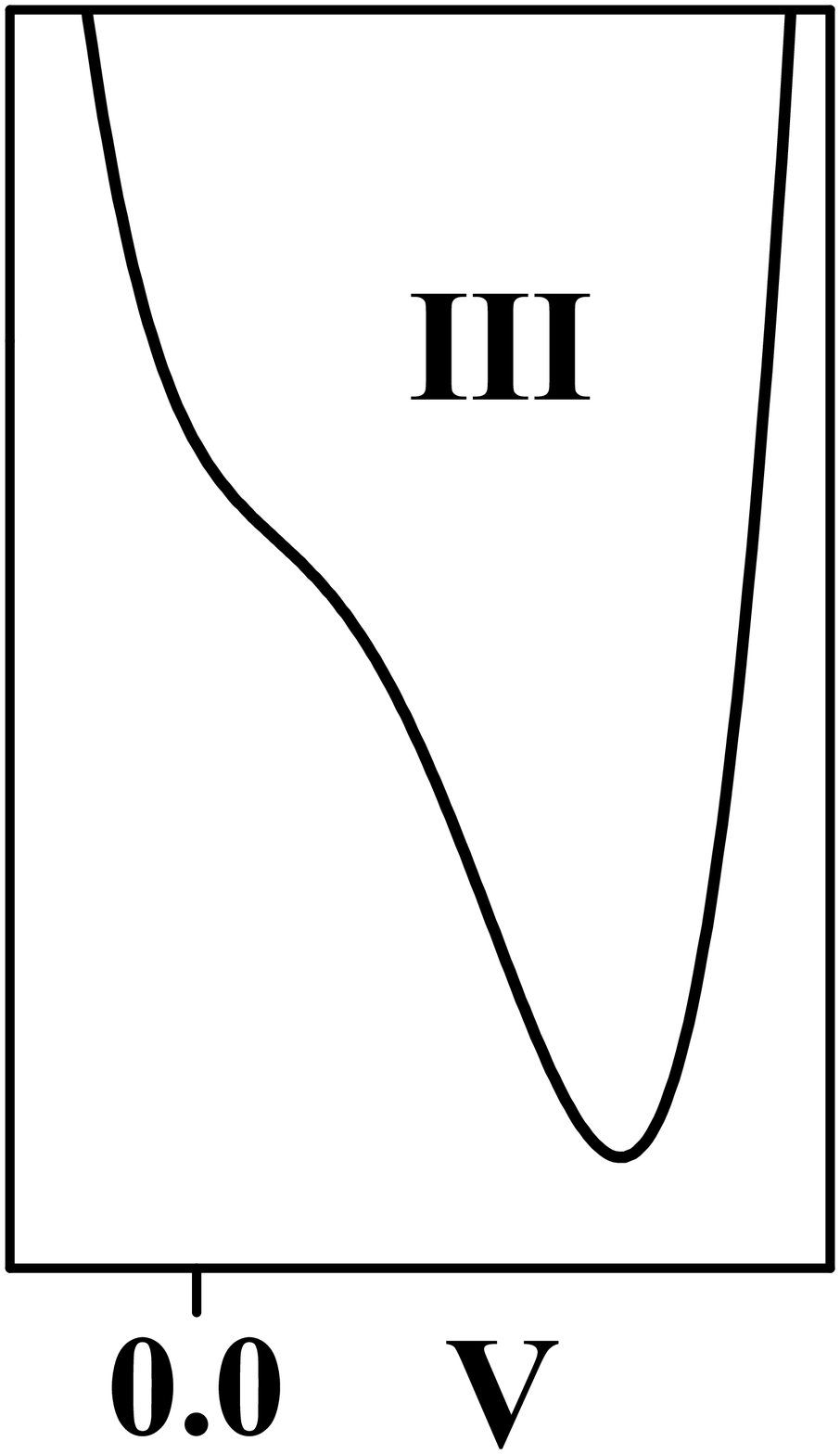} $\!\!\!\!\!$
       \caption{(Color online) Dependence of the dimensionless diffusivity on
       the external constant forcing at different temperatures (top).
       The region of anomalous temperature dependence (II) is hatched. The
       friction coefficient $\gamma^\prime=0.2$. 1: $T^\dagger =0.097$
       [$T = 90\:$K for the potential characteristic of the H-adatom 
	   diffusion on metal surface (\ref{U0a})], 2: $T^\dagger = 0.129$, 3: $T^\dagger = 0.194$,
       4: $T^\dagger = 0.388$, 5: $T^\dagger = 0.582$, 6: $T^\dagger = 0.776$,
       7: $T^\dagger = 0.969$.\\
       Bottom: the effective velocity potential in the three force regions.
       Left, middle and right diagrams correspond to $F_e/F_0=0.06$, 0.095 and
       0.25 respectively.}
    \label{f:difconstF}
  \end{center}
\end{figure}

Plots of the diffusivity as a function of the external force at several
temperatures (based on \cite{Marchenko12EPL}, recomputed for the current units)
are shown in Fig.~\ref{f:difconstF}. There are three regions of $F_e$ values
(I--III), with qualitatively different dependence of the diffusivity $D$ on
$T$.
In region I diffusion is enhanced as external force $F_e$ value increases,
whereas in region III diffusion is inhibited as $F_e$ increases. In regions
I and III $D$ grows with $T$ increasing. Contrary to this (normal temperature
behavior) in region II $D$ grows as $T$ decreases. This is the region
of ``temperature-abnormal diffusivity'' (TAD). The \emph{diffusion process}
at sufficiently late times is normal, $\sigma^2\propto t^1$, however the
\emph{diffusion coefficient} $D^\prime=\sigma^2/(2t^\prime)$ (abnormally)
increases as the temperature decreases.

In \cite{Marchenko12JETPL} it was seen that TAD is also possible at
time-periodic external forcing. At such forcing TAD was only observed in
a limited interval of the forcing frequencies. Studies were carried out at
$F_e=0.15F_0$, that would correspond to region III for the same amplitude of
constant forcing (see Fig.~\ref{f:difconstF}), so TAD would not be observed at
such $F_e$ value were the forcing $F_t(t)$ constant.

To understand physical reasons of the abnormal temperature dependence of
the diffusivity in these two different (time-periodic, and constant) forcing
regimes we consider the periodic forcing regime in more detail below. For
piecewise constant periodic (PCP) forcing one can use intuition gained in the
constant forcing problem for the time intervals in which $F_t$ stays constant,
the findings are thus easier to interpret. Most of the results in this work
are for the PCP forcing setup.


In \cite{MarchenkoFconst} we showed that the transport
properties of an ensemble of underdamped particles in a washboard potential
can be found by considering a simpler problem of the overdamped motion in
effective potential $W(V)$ in velocity space. The problem showed
similarities with the problem of active Brownian particle motion
\cite{Lindner.GiantDifsn}. Whether the particle belongs to locked
[tending with time to $V=0$, one minimum of $W(V)$] or running [tending to
$V=F/\gamma$, another minimum of the asymmetric double-well potential $W(V)$]
population in the absence of noise is determined by the particle's initial
state (its velocity, if placed at the potential minimum at $t = 0$). It was
shown that region II of the force values, the region in which TAD is observed,
is precisely the region where both locked and running solutions are
simultaneously possible.

The $W(V)$ was also instrumental for proposing a model distribution function
$N(V)$, and for defining diffusivity through Kubo relation. The analytic
predictions for $D$ and $\langle V\rangle$ based on the model $N(V)$ agreed
well with direct numerical simulation results for all values of $F\lesssim 0.5
F_0$, temperatures in the range $T^\dagger\in[0.129;0.776]$, friction
$\gamma^\prime\in[0.1;0.4]$ (in the definitions of dimensionless quantities
adopted here, Eqs.~(\ref{dimtx},\ref{dimTgamma}); agreement getting
progressively better at smaller $\gamma^\prime$). In the expression for
the diffusivity, (the analytic prediction for) the correlation time
$\tau_{corl}$ factor was shown to be responsible for the main features in
$D(F,T)$ dependence, numerically found in earlier study \cite{Marchenko12EPL}.
Namely, $\tau_{corl}$ showed proper nonmonotonic dependence on $F$, with a
maximum near $F_\mathrm{cr}$, progressively more pronounced at smaller $T$.
$\tau_{corl}$ increased at temperature decreasing
$\propto \exp[\mathcal{E}/(k_BT)]$ (for certain $\mathcal{E}>0$
); this increase at low $T$
reflected the Kramers rate for transitions in the double-well
velocity potential $W(V)$, between the wells corresponding to locked and
running states.

Schematics of the effective velocity potential in the three regions
of the applied force are shown in the bottom insets in Fig.~\ref{f:difconstF}.
The first inset corresponds to $F_e=0.06F_0$, at which only locked
solutions exist. In the second inset $F_e=0.095F_0$, corresponding to the
maximal diffusivity; both types of solutions coexist at this force value.
Finally $F_e=0.25F_0$ corresponding to region III is shown in the third inset;
at this $F_e$ only running solutions are realized. The boundaries between
regions I--III depend on $\gamma$ and $m$ (on $\gamma^\prime$ in the
dimensionless problem). At smaller $\gamma^\prime$ these boundaries move to
the left, to the lower force values, and the width of region II decreases.
In the current study we use the same fixed values of $\gamma$ and $m$ as in
\cite{MarchenkoFconst}, and investigate the features of $D(T|\omega)$ at the
three values of the external forcing amplitude listed above.

It may be expected that the same picture for different temperature behaviors
of the diffusivity, with analogous three regions I--III of the external force
amplitudes, holds for PCP driving, at least at small frequencies $\Omega$.
The description with the effective potential is applicable, the
potential $W(V)$ keeping constant form for each of the two half-periods of the
external forcing [on which $F_t$ stays constant, this $F$ value defining
corresponding $W(V|F,\gamma)$.]
%
%

Below we show that in fact the situation is more subtle.
Similar three regions of the external force amplitudes $F_e$ exist, differing
in qualitative $D(T)$ dependence. TAD is realized in certain range of
intermediate $F_e$, smoothly dependent on $\Omega$. At decreasing
temperatures the dependence $D(T)$ starts progressively
deviating from such at constant forcing. For any fixed $\Omega$ at temperatures
below certain critical $T_\textrm{TAD}(\Omega)$, normal temperature dependence
of the diffusivity is restored.
We demonstrate these features by analyzing $D(T|\omega)$ in sequence for the
above values of $F_e$ ($F_e/F_0=0.06$, 0.095, 0.25), that correspond to regions
I, II, III in the constant external force problem.

We observe transient anomalous diffusion phase, and see how its characteristics
relate to the diffusivity in the asymptotic late time regime, when normal
diffusion regime sets in. The properties of this anomalous diffusion stage
are different in the three regions of the applied force amplitudes; this
translates into different functional dependence of the diffusivity on the
temperature.
\subsection{Region I:  $F_e/F_0=0.06$, varying frequency}\label{s:RegionI}
Here we investigate how the diffusivity changes with (dimensionless) frequency
$\omega=\Omega\mathcal{T}_0$ of the external forcing at forcing amplitude
$F_e=0.06F_0$. We show that the diffusivity stays nearly constant at small
frequencies, up to reciprocal of the superdiffusion stage duration. At
frequencies above this the diffusivity drops according to a power-law with
exponent related to the exponent of the superdiffusion. This power-law
dependence holds up to frequencies a few times less than the lattice potential
eigenfrequency. The diffusivity grows with the temperature (normal temperature
behavior), related to the superdiffusion stage duration increasing with $T$.

\begin{figure}[htbp!]
  \begin{center}
    \includegraphics[width=0.47\textwidth]{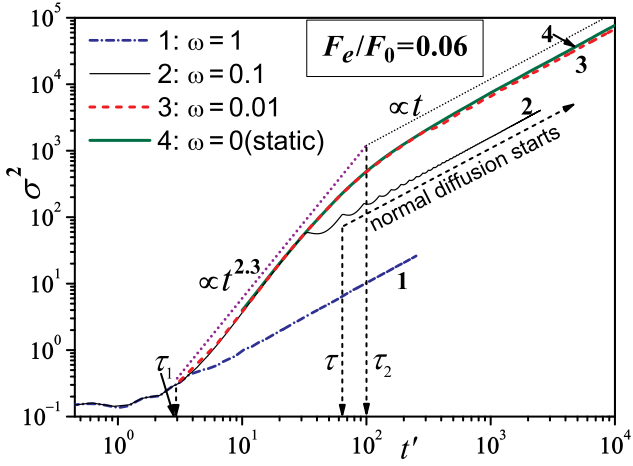}
       \caption{(Color online) Time dependence of the dimensionless dispersion
       for the force value $F_e=0.06 F_0$ (region I) for different force
       frequencies. $\tau_2$ denotes the end of the superdiffusion stage at
       constant forcing. $\tau=2\pi/\omega$ is the period of the external
       periodic forcing. $T^\dagger=0.194$.
       1: $\omega=1$, 2: $\omega=10^{-1}$, 3: $\omega=10^{-2}$,
       4: $\omega=0$ (constant external force.) Straight dashed and dotted
       lines show the power-law fitting behavior of the dispersion.}
    \label{f:sigma006}
  \end{center}
\end{figure}

Figure~\ref{f:sigma006} shows the growth of the particle dispersion with time.
At constant external forcing this growth demonstrates a clear superdiffusion
stage, $\sigma^2\propto t^{\prime\alpha}$, $\alpha>1$, that switches to normal
diffusion (same power-law growth but with $\alpha =1$) after certain time $\tau_2$.
The $\tau_2$ as well as the superdiffusion exponent $\alpha$ depend on
$F_e/F_0$ and other problem parameters, $\gamma^\prime$ and $T^\prime$.
For the values used $\tau_2\approx 100$, $\alpha\approx 2.3$. Lowercase
$\tau$'s everywhere are dimensionless, normalized to the period of small
oscillations at the lattice potential minimum at $\gamma^\prime=0$,
\begin{equation}\label{tau0}
  \mathcal{T}_0=2\pi/\Omega_0=a\sqrt{2m/U_0}.
\end{equation}
$\sigma^2$ and $D^\prime$ are dimensionless dispersion and diffusivity.

Curves 1--3 correspond to three different frequencies. Oscillations in
$\sigma^2(t^\prime)$ are clearly seen with double forcing frequency, $2\omega$.
These are particularly large at $\omega$ around
$\omega_2=2\pi/\tau_2$ (curve 2). At these values, interestingly, intervals of
$\sigma^2(t)$ decreasing with time are observed. This effect can be
seen in earlier works \cite{SaikiaMahato09PRE}.

At frequencies below
$\omega_2=2\pi/\tau_2$ the curves $\sigma^2(t)$ closely follow that at
$\omega=0$ (constant external force, curve 4). At higher frequencies the
superdiffusion phase
effectively ends after one period $\tau$ of the external force, normal
diffusion sets in (in a sense that averaged over the forcing period
$\langle d(\sigma^2)/dt\rangle_t \approx const$).
Higher frequencies
of the forcing correspondingly result in lower diffusivity, as the stage of
$\sigma^2$ growing with time faster than linearly ends earlier than at lower
$\omega$'s.

These results suggest the following approximations for the smoothed
over the forcing period dispersion
$\overbar{\sigma^2}(t^\prime)$, valid not too close to the transition moments
between different diffusion regimes:
\begin{equation}\label{sigma2fitI}
  \ln{\frac{\overbar{\sigma^2}}{\sigma^2_0}}=\left\{  \begin{array}{ll}
  \approx 0 & \textrm{at }t^\prime<\tau_1 \\
  \alpha\ln(t^\prime/\tau_1) &
     \textrm{at } t^\prime\in(\tau_1;\tau_{2,\omega}) \\
  \alpha\ln(\tau_{2,\omega}/\tau_1)+\ln(t^\prime/\tau_{2,\omega}) &
     \textrm{at }t^\prime>\tau_{2,\omega}.
  \end{array}\right.
\end{equation}
Here $\sigma^2_0=\sigma^2|_{t=0}$ (for the approximately Boltzmannian initial
distribution of the particles in $x^\prime\in(-0.5;0.5)$ well of the
potential), $\tau_{2,\omega}=\mathrm{min}(\tau_2,\tau)$, $\tau_1$ is close
to the first potential well exit time. Constants
are chosen in such a way that the spline approximation (\ref{sigma2fitI}) is
continuous. Quantities $\alpha$, $\tau_{1,2}$ depend on the problem parameters,
$T^\prime$, $F_e/F_0$ and $\gamma^\prime$. At fixed values of these parameters,
the last approximation yields the dependence of the diffusivity
$D^\prime=\lim\limits_{t^\prime\to\infty}\sigma^2/(2t^\prime)$ on the forcing
period $\tau$
\begin{equation}\label{DtaufitI}
  D^\prime(\tau)=\left\{  \begin{array}{ll}
  (1/2)\sigma^2_0\tau_1^{-\alpha}\tau^{\alpha-1} & \textrm{at }\tau_1<\tau<\tau_2 \\
  (1/2)\sigma^2_0\tau_1^{-\alpha}\tau_2^{\alpha-1} & \textrm{at }\tau>\tau_2,
  \end{array}\right.
\end{equation}
or on its frequency $\omega=2\pi/\tau$
\begin{equation}\label{DomegafitI}
  \frac{D^\prime(\omega)}{D^\prime(0)}=\left\{  \begin{array}{ll}
  1 & \textrm{at }\omega<\omega_2 \\
  (\omega/\omega_2)^{1-\alpha} & \textrm{at }\omega_2<\omega<2\pi/\tau_1.
  \end{array}\right.
\end{equation}
The diffusivity thus should stay about constant at
$\omega<\omega_2$, and at larger forcing frequencies drop
with $\omega$ according to the power law, with exponent equal to $1-\alpha$,
determined by the superdiffusion exponent.

\begin{figure}[htbp!]
  \begin{center}
    \includegraphics[width=0.47\textwidth]{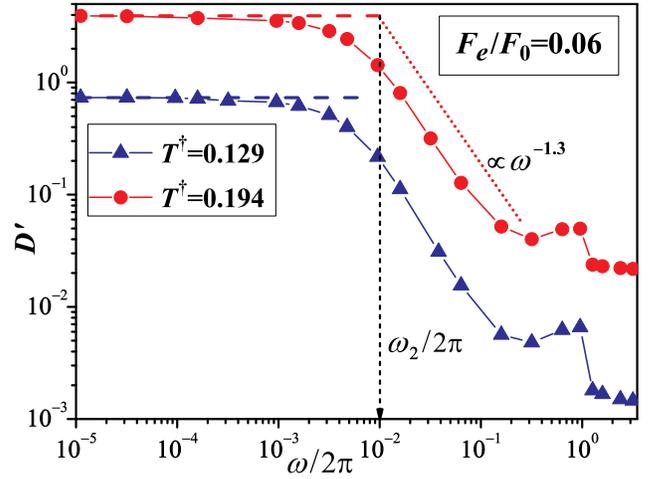}
       \caption{(Color online) Dependence of diffusivity
       $D^\prime$ on the forcing frequency at
       $T^\dagger=0.129$ (triangles) and $0.194$ (circles). The dashed line
       asymptotes show the diffusivity values at constant forcing with the same
       $F_e/F_0=0.06$. The slanting dotted line shows the power law drop of
       $D^\prime$ at intermediate $\omega$'s. }
    \label{f:Domega006}
  \end{center}
\end{figure}

Figure~\ref{f:Domega006} shows simulation results for $D^\prime(\omega)$.
Dotted lines show approximations Eq.~\ref{DomegafitI}. Dashed lines show the
diffusivity at constant forcing with the same $F_e$ value. These results
confirm the features suggested by Eq.~\ref{DomegafitI}, away from
transition frequency $\omega_2=2\pi/\tau_2$ and up to $\omega/2\pi\approx 0.1$.
$D(\omega)<D(0)$ at any $\omega$, and decreases with $\omega$, up to the
frequency values a few times smaller than the potential eigenfrequency.

Further ramification might have been added in
Eqs.~(\ref{sigma2fitI}--\ref{DomegafitI}) for the case $\tau<\tau_1$, changing
$\tau_1$ to $\tau_{1,\omega}=\mathrm{min}(\tau_1,\tau)$; that would indeed
lead to $D(\omega)$ increasing at $\omega>2\pi/\tau_1$. With our parameters,
however, resulting approximations are an overstretch; the diffusion gets
markedly nonequilibrium at such high $\omega$'s, resulting in the complex
$D(\omega)$ behavior at $\omega$ approaching $2\pi$. More on this in
Sec.~\ref{s:HighOmega}.

\begin{figure}[htbp!]
  \begin{center}
    \includegraphics[width=0.47\textwidth]{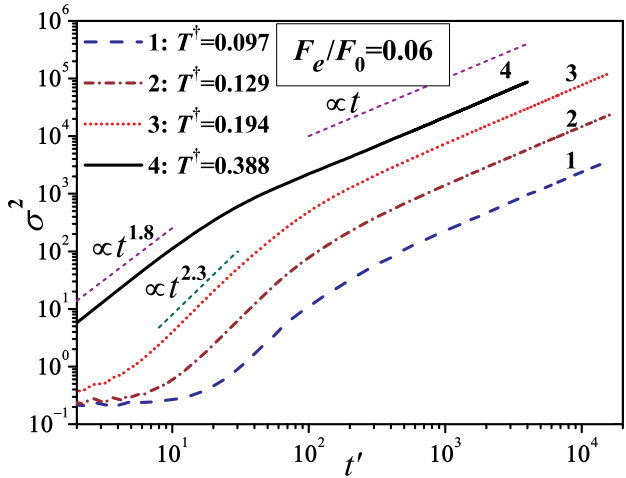}
       \caption{(Color online) Dispersion variation with the temperature, at
       constant forcing ($\omega=0$). 1: $T^\dagger=0.097$, 2:
       $T^\dagger=0.129$, 3: $T^\dagger=0.194$, 4:  $T^\dagger=0.388$.
       The short-dashed straight lines show power-law fits for the transient
       superdiffusion stage, and the late-time normal diffusion regime.}
    \label{f:sigmaT006}
  \end{center}
\end{figure}

Finally, Fig.~\ref{f:sigmaT006} shows $\sigma^2(t)$  for a range of
temperatures at constant forcing. Despite the superdiffusion exponent $\alpha$
is seen to decrease with the temperature, the duration of the superdiffusion
stage $\tau_\textrm{spd}=\tau_2-\tau_1$ grows fast with the temperature. As a
result the dispersion is larger for larger temperatures at any given time. At
late times this translates into diffusivity being larger at larger
temperatures; so normal (increasing) $D(T)$ dependence is realized at force
amplitudes in region I.
\subsection{Region II: $F_e/F_0=0.095$}\label{s:RegionII}
Here we repeat analysis of the previous section for periodic forcing with
amplitude $F_e=0.095F_0$ (the value at which the maximal diffusivity is
achieved at constant forcing). In the same manner as for the forcing amplitude
in region I, transition from initial superdiffusion stage to normal diffusion
at late times
is observed, at time $\tau_2$. At forcing frequency $\omega$ growing,
the diffusivity again decreases only slightly from $D(\omega=0)$ till $\omega$
reaches $\omega_2 = 2\pi/\tau_2$. It decreases fast, according to the same
power law $D(\omega)/D(0)=(\omega/\omega_2)^{1-\alpha}$, at larger $\omega$'s,
up to $1/\tau\approx 0.1$. Contrary to the case of weaker forcing considered in
the previous section, superdiffusion duration $\tau_\textrm{spd}=\tau_2-\tau_1$
now \emph{decreases} with the temperature. This leads to diffusivity growing at
temperature decrease (TAD), at moderately low frequencies. This is the main
effect we intended to find in the periodic driving setup, extending our
previous findings at constant forcing \cite{MarchenkoFconst}.

\begin{figure}[htbp!]
  \begin{center}
    \includegraphics[width=0.47\textwidth]{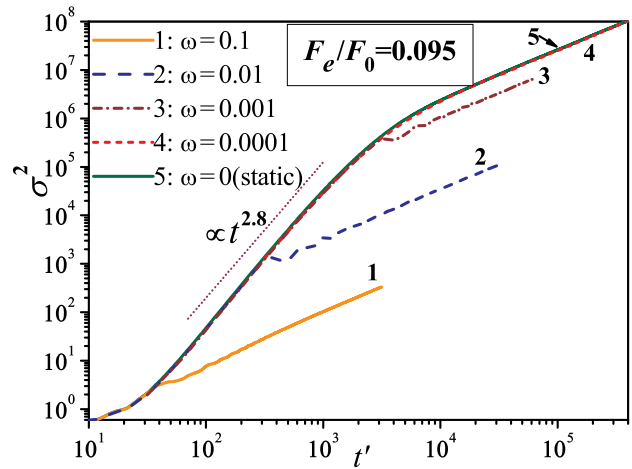}
       \caption{(Color online) Time dependence of the dimensionless dispersion
       for the force value $F_e=0.095F_0$ (region II) for different force
       frequencies. Curves 1--5, bottom to top: (dimensionless) $\omega=$
       $10^{-1}$, $10^{-2}$, $10^{-3}$, $10^{-4}$, 0 (constant forcing).
       $T^\dagger=0.194$. }
    \label{f:sigma0095}
  \end{center}
\end{figure}

\begin{figure}[htbp!]
  \begin{center}
    \includegraphics[width=0.47\textwidth]{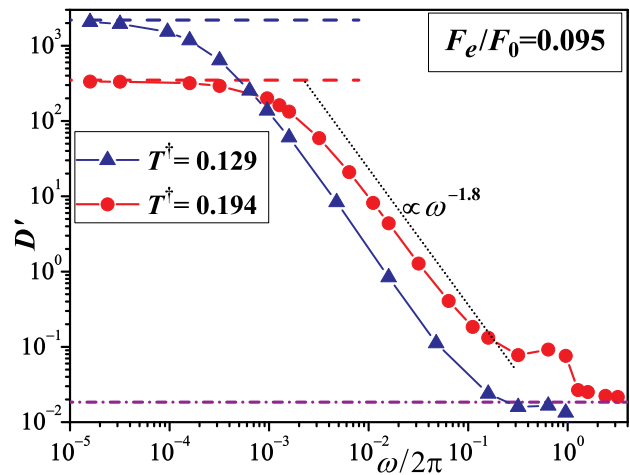}
       \caption{(Color online) Dependence of diffusivity $D^\prime$ on the
       external force frequency $\omega$, at $T^\dagger=0.129$ (triangles)
       and $0.194$ (circles). The dashed line asymptotes show the diffusivity
       values for constant forcing with the same $F_e/F_0=0.095$. Dotted
       line shows the power law drop of $D^\prime$ at higher $\omega$'s.
       The dot-dashed horizontal line asymptote (for high $\omega$) is at the
       diffusivity value in the lattice periodic potential in the absence
       of extra driving, $F_t(t)=0$, for $T^\dagger=0.194$.}
    \label{f:Domega0095}
  \end{center}
\end{figure}

Figure~\ref{f:sigma0095} shows growth of the particle dispersion with time.
Again the superdiffusion stage is observed, and it ends after one period of
PCP forcing or after $\tau_2$ (the end time of superdiffusion  at constant
forcing), whichever is smaller. Thus the same approximations
Eq.~(\ref{sigma2fitI}--\ref{DomegafitI}) are applicable for the (averaged over
$\tau$) dispersion
growth $\overbar{\sigma^2}(t^\prime)$ and dependence of the diffusivity
on the forcing
frequency $D^\prime(\omega)$. Simulation results for $D^\prime(\omega)$
(Fig.~\ref{f:Domega0095}) again agree with prediction (Eq.~\ref{DomegafitI}).

As $D(\omega)$ does not vary much at $\omega<\omega_2$, TAD must be observed at
small frequencies, at it is observed at constant forcing
(\cite{MarchenkoFconst}: $D(F_e,\omega=0)$ decreases with the temperature
when $F_e$ is in region II). Such an abnormal dependence on the temperature is
due to longer jumps the particles perform on average at lower temperatures.
This in turn follows from lower probability at lower temperatures to turn the
running particle into a locked state: such a transition requires certain
threshold energy transfer to such a particle, and at lower temperatures the
probability of such a transfer decreases exponentially with inverse
temperature. Hence the running particle travels at about the stationary speed
$F_e/\gamma$ (the speed fluctuating
around this average due to the lattice potential) for increasingly longer
distances before getting trapped again in some next well of the potential,
resulting in larger diffusivity at smaller
temperatures. This mechanism is not applicable in region I, where only
locked solutions are realized.

\begin{figure}[htbp!]
  \begin{center}
    \includegraphics[width=0.47\textwidth]{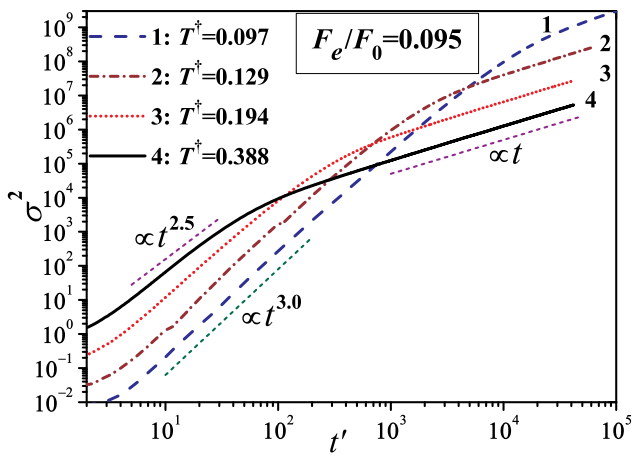}
       \caption{(Color online) The dispersion variation with the temperature,
       at constant forcing ($\omega=0$).
       1: $T^\dagger=0.097$, 2: $T^\dagger=0.129$, 3: $T^\dagger=0.194$, 4:
       $T^\dagger=0.388$. $F_e/F_0=0.095$.}
    \label{f:sigmaT0095}
  \end{center}
\end{figure}

In terms of the dispersion evolution features, this phenomenon of the
diffusivity growing at the temperature decreasing
results from the superdiffusion duration $\tau_\mathrm{spd}=\tau_2-\tau_1$
growing at temperature decreasing, as Fig.~\ref{f:sigmaT0095} demonstrates.
This qualitatively differs from the situation at forcing amplitude in region I
(end of the previous section). We note that $\alpha$ increases with the
temperature decreasing, same as in region I.

At higher frequencies, on the other hand, the diffusivity starts increasing
with the temperature (normal behavior). This results from $\tau_2$ decreasing
with the temperature, as can be seen in Fig.~\ref{f:sigmaT0095}. Despite
$D(\omega=0)$ is larger at smaller temperature $T_1$, as $\omega$ grows
the $D(\omega\,| T_1)$ starts decreasing according to the power law earlier
[at $\omega$'s near the smaller $\omega_2=2\pi/{\tau_2(T_1)}$] than
$D(\omega\,| T_2)$ for $T_2> T_1$ does. And the exponent of this power-law
decreases with $T$ decreasing,
$1-\alpha(T_1)< 1-\alpha(T_2)<0$ --- therefore at larger frequencies,
commensurate with $\omega_2$, the diffusivity starts increasing with the
temperature, \emph{TAD disappears}.

As it is the case for the forcing amplitude in region I, $D(\omega)$ decreases
monotonically with $\omega$ from $D(0)$, up to the frequency values
$\omega\approx 0.3\times 2\pi$ (at frequencies so high the particle transport becomes
substantially non-equilibrium, that affecting the functional dependence $D(\omega)$,
as discussed in Sec.~\ref{s:HighOmega} below).
\subsection{Region III: $F_e/F_0=0.25$}\label{s:RegionIII}
For strong forcing (region III, in which the diffusivity decreases with $F_e$
in constant forcing setup, and increases with the temperature) we demonstrate
in this subsection that the dependence of dispersion on time shows new
features. After
initial superdiffusion stage at times $t^\prime<\tau_2$, prolonged
dispersionless stage sets in, till $\tau_3\sim 10^3 \tau_2$. In this stage
the dispersion stays nearly constant. As at weaker external forcing
(regions I, II), both these stages may be interrupted earlier in the case of
high forcing frequency (after about a period of the forcing). At late times
normal diffusion takes place. Dependence $D(\omega)$ is nonmonotonic here,
agreeing with findings in \cite{Marchenko12JETPL}. $D$ reaches a maximum near
$\omega=\omega_2\equiv 2\pi/\tau_2$. TAD is observed at intermediate
frequencies.

\begin{figure}[htbp!]
  \begin{center}
    \includegraphics[width=0.47\textwidth]{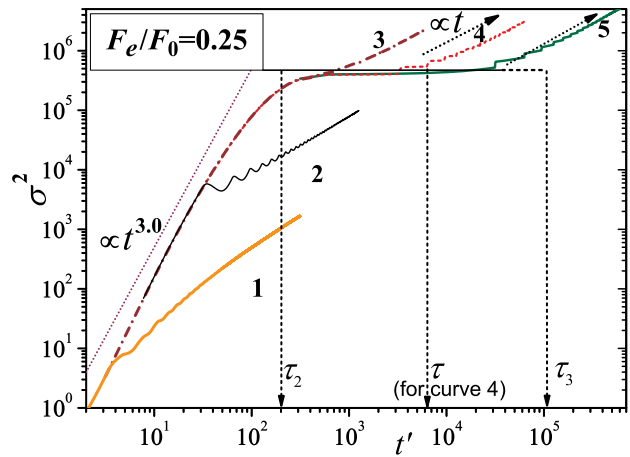}
       \caption{(Color online) Time dependence of the dimensionless dispersion
       at $F_e=0.25F_0$ (region III) for different force frequencies.
       Curves 1--5: (dimensionless) $\omega=$ $1$, $10^{-1}$, $10^{-2}$,
       $10^{-3}$, $10^{-4}$. $T^\dagger=0.194$.
       Dotted lines show power-law fitting behavior of the
       dispersion.}
    \label{f:sigma025}
  \end{center}
\end{figure}

Figure~\ref{f:sigma025} shows the particle dispersion as a function of time.
At low frequencies of the forcing, superdiffusion is observed till
$\tau_2\approx 200$. Dispersionless stage is realized later, at
$t^\prime\in(\tau_2; \tau_3)$. Physically $\tau_2$ is the time at which the
particle distribution function in velocities assumes its 
stationary form
. The distribution in
space is still strongly non-equilibrium at $\tau_2$, it has exponential tail
and sharp front \cite{Lindenberg07PRL} (in the direction of $F$). It takes till
$\tau_3$ for the
distribution function to assume approximately Gaussian shape in space (if
averaged over the potential spatial period).
On interval $(\tau_2; \tau_3)$ dispersion does not increase siubstantially.
After $\tau_3$ normal diffusion sets in, $\sigma^2/t^\prime=2 D^\prime+o(1)$
at $\tau_3\ll t^\prime\to\infty$.

The whole curve $\sigma^2(t^\prime)$ differs insignificantly from
$\sigma^2(t^\prime)$ in constant forcing setup if the forcing frequency
$\omega<\omega_3\equiv 2\pi/\tau_3$. This $\sigma^2(t^\prime)$ dependence is
modified at higher frequencies of the external forcing (in a similar way to
the behavior discussed in Sec.~\ref{s:RegionI}--\ref{s:RegionII} at weaker
forcing). For $\omega\in (\omega_3; \omega_2)$ the dispersionless stage is
interrupted after about one period of external force, after which time
approximately linear dependence (if averaged over times $\sim 2\pi/\omega$)
sets in, $\langle d\sigma^2/dt^\prime\rangle_{t^\prime}=2D^\prime$, $D^\prime$
equal to its asymptotic (at $t^\prime\to\infty$) value. At yet higher forcing
frequencies, $\omega >\omega_2$, dispersionless stage is not realized at all,
normal diffusion sets in right after the prematurely interrupted superdiffusion
stage.

Similarly to the previous sections, described above dependence of
$\ln\overbar{\sigma^2}$ on
$\ln t^\prime$ can be schematized by the following continuous piecewise linear
approximation (meaningful not too close to switching time-points $\tau_{1-3}$):
\begin{equation}\label{sigma2fitIII}
  \ln{\frac{\overbar{\sigma^2}}{\sigma^2_0}}=\left\{  \begin{array}{ll}
  \approx 0 & \textrm{at }t^\prime<\tau_1 \\
  \alpha\ln(t^\prime/\tau_1) &
       \textrm{at } t^\prime\in(\tau_1;\tau_{2,\omega}) \\
  \alpha\ln(\tau_{2,\omega}/\tau_1)   &
       \textrm{at } t^\prime\in(\tau_{2,\omega};\tau_{3,\omega}) \\
  \alpha\ln(\tau_{2,\omega}/\tau_1)+\ln(t^\prime/\tau_{3,\omega}) &
       \textrm{at }t^\prime>\tau_{3,\omega}.
  \end{array}\right.
\end{equation}
Here $\tau_{3,\omega}=\mathrm{min}(\tau_3,\tau)$,
$\tau_{2,\omega}=\mathrm{min}(\tau_2,\tau)$; $\tau>\tau_1$ assumed.
As before, $D(\omega)$ can be found from this, showing the same power law
decay as in Eq.~\ref{DomegafitI} at $\omega>\omega_2$, and additional,
linearly growing with $\omega$ feature at intermediate frequencies:
\begin{equation}\label{DomegafitIII}
  \frac{D(\omega)}{D(0)}=\left\{  \begin{array}{ll}
  1 & \textrm{at }\omega<\omega_3 \\
  \omega/\omega_3 & \textrm{at }\omega\in(\omega_3;\omega_2) \\
  (\omega/\omega_2)^{1-\alpha}\omega_2/\omega_3 & \textrm{at }\omega>\omega_2.
  \end{array}\right.
\end{equation}

\begin{figure}[htbp!]
  \begin{center}
    \includegraphics[width=0.47\textwidth]{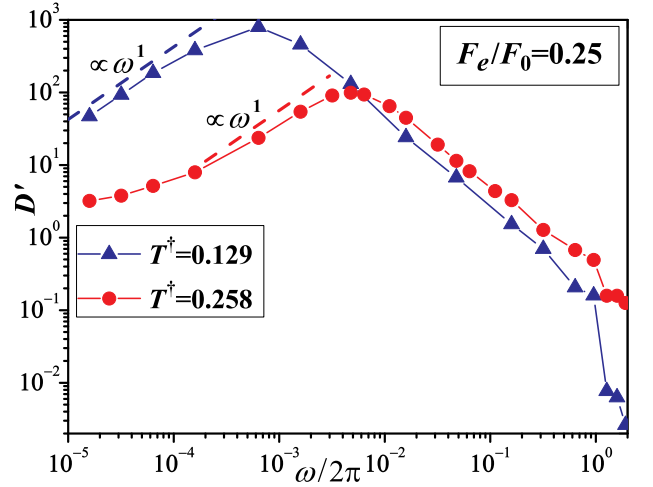}
       \caption{(Color online) Dependence of the dimensionless diffusivity
       $D^\prime$ on the forcing frequency, at $T^\dagger=0.129$ (circles)
       and 0.194 (triangles). The dashed lines show approximately linear
       growth at $\omega\in(\omega_2; \omega_3)$.}
    \label{f:Domega025}
  \end{center}
\end{figure}

Simulation results for $D(\omega)$ are shown in Fig.~\ref{f:Domega025} for two
temperatures. They indeed show the features suggested in
Eq.~\ref{DomegafitIII}. At high frequencies, $\omega>\omega_2$, the
superdiffusion stage switches to normal diffusion earlier than at $\tau_2$,
at time $\sim \tau=2\pi/\omega$. This behavior is the same as for the forcing
amplitude in
regions I--II. Thus in the same manner the diffusivity grows as $\omega$
decreases, as the longer interval (till $t^\prime\approx\tau$) of
superdiffusive dispersion growth is used at the forcing period $\tau$
getting longer.
However at smaller frequencies,
$\omega_3<\omega<\omega_2$, it is the dispersionless phase that gets 
interrupted prematurely (\emph{i.e.} it is not realized in full, till $\tau_3$, 
as it would have at $\omega=0$),
so the earlier such a switch occurs (\emph{i.e.} the shorter the $\tau$),
the higher
diffusivity is obtained in the late-time normal diffusion stage. This explains
the growth of $D(\omega)$ with $\omega$ at $\omega\in(\omega_3 ;\omega_2)$.
The maximum in $D(\omega)$ is thus observed, at
$\omega_\textrm{max}\approx \omega_2$.

This differs from monotonically decreasing $D(\omega)$ in regions I, II. Such a
nonmonotonic dependence $D(\omega)$ was observed in
\cite{Marchenko12JETPL,Marchenko14JPhCS};
the forcing studied was sinusoidal in time, its amplitude (indeed) corresponded
to region III in terminology of this work.

{$\omega_\textrm{max}$ together with $2\pi/\tau_2$ increase with
the temperature.
The maximal value of the diffusivity (achieved at $\omega_\textrm{max}$)
increases as the temperature decreases, due to the dispersionless stage
duration $\tau_3-\tau_2$ growing fast at temperature decreasing.
And so does $D(\omega)$ at a given $\omega$
near $\omega_2$ in a limited range of temperatures. So abnormal temperature
dependence of the diffusivity (TAD)
is realized in intermediate region of the forcing
frequencies around $\omega_2$. The temperature dependence of the
diffusivity is studied in more detail in Sec.~\ref{s:DvsT} below.}
\subsection{Diffusivity at high forcing frequency $\omega$}\label{s:HighOmega}
As the simulation results suggest, the power law decay of $D(\omega)$ at large
$\omega$'s (Figs.~\ref{f:Domega006}, \ref{f:Domega0095}, \ref{f:Domega025})
does not hold at frequencies that are too high, above $\sim 1/10^\textrm{th}$
of the eigenfrequency at the lattice potential minima. This is due to the
diffusion process becoming markedly non-equilibrium.  The distribution function
in velocities differs significantly from such at constant forcing at all times;
description in terms of the velocity potential $W(V)$ is no longer useful.

\begin{figure}[htbp!]
  \begin{center}
    $\!\!\!$\includegraphics[width=0.241354\textwidth]{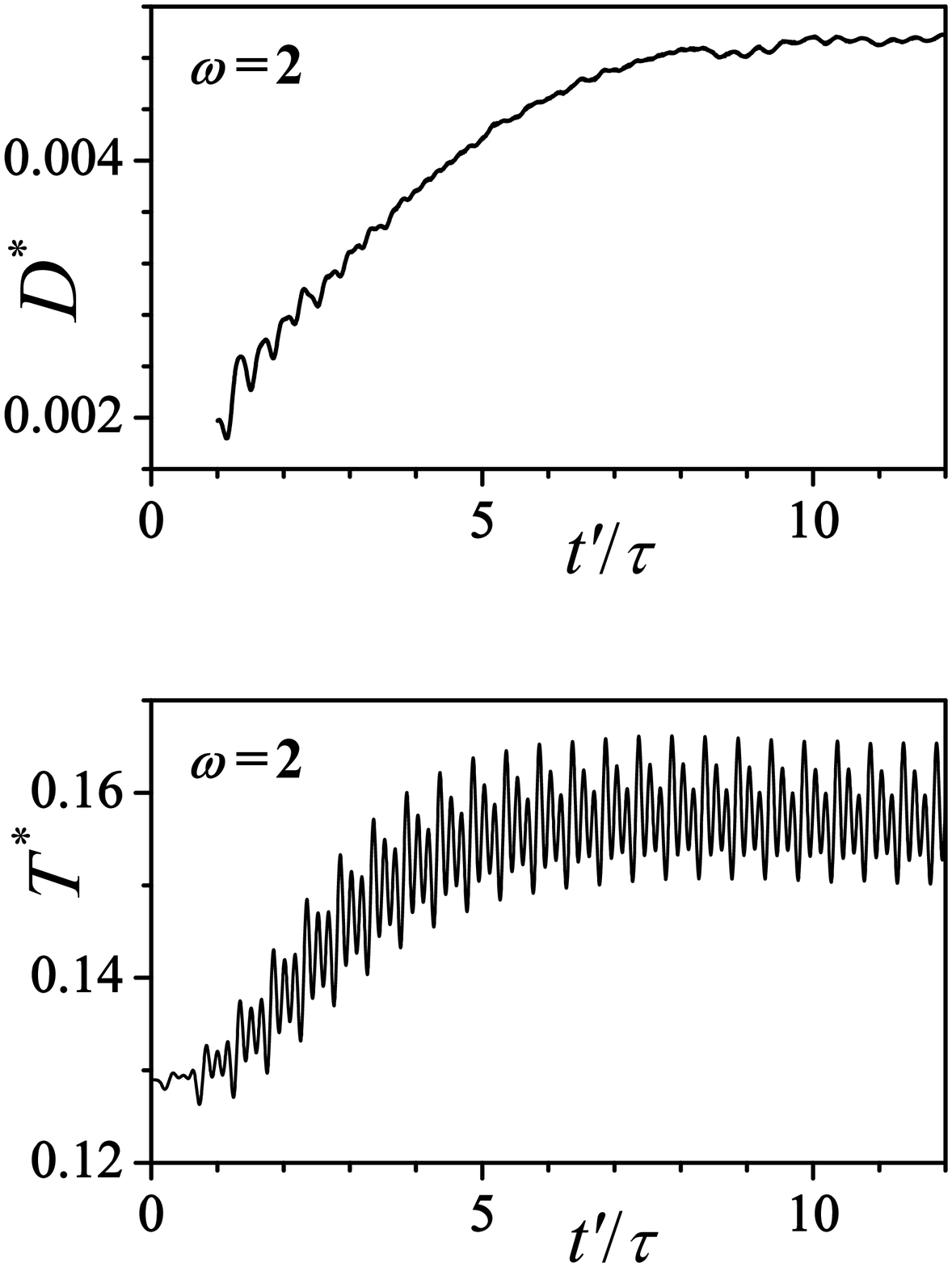}
    $\!\!\!$\includegraphics[width=0.24\textwidth]{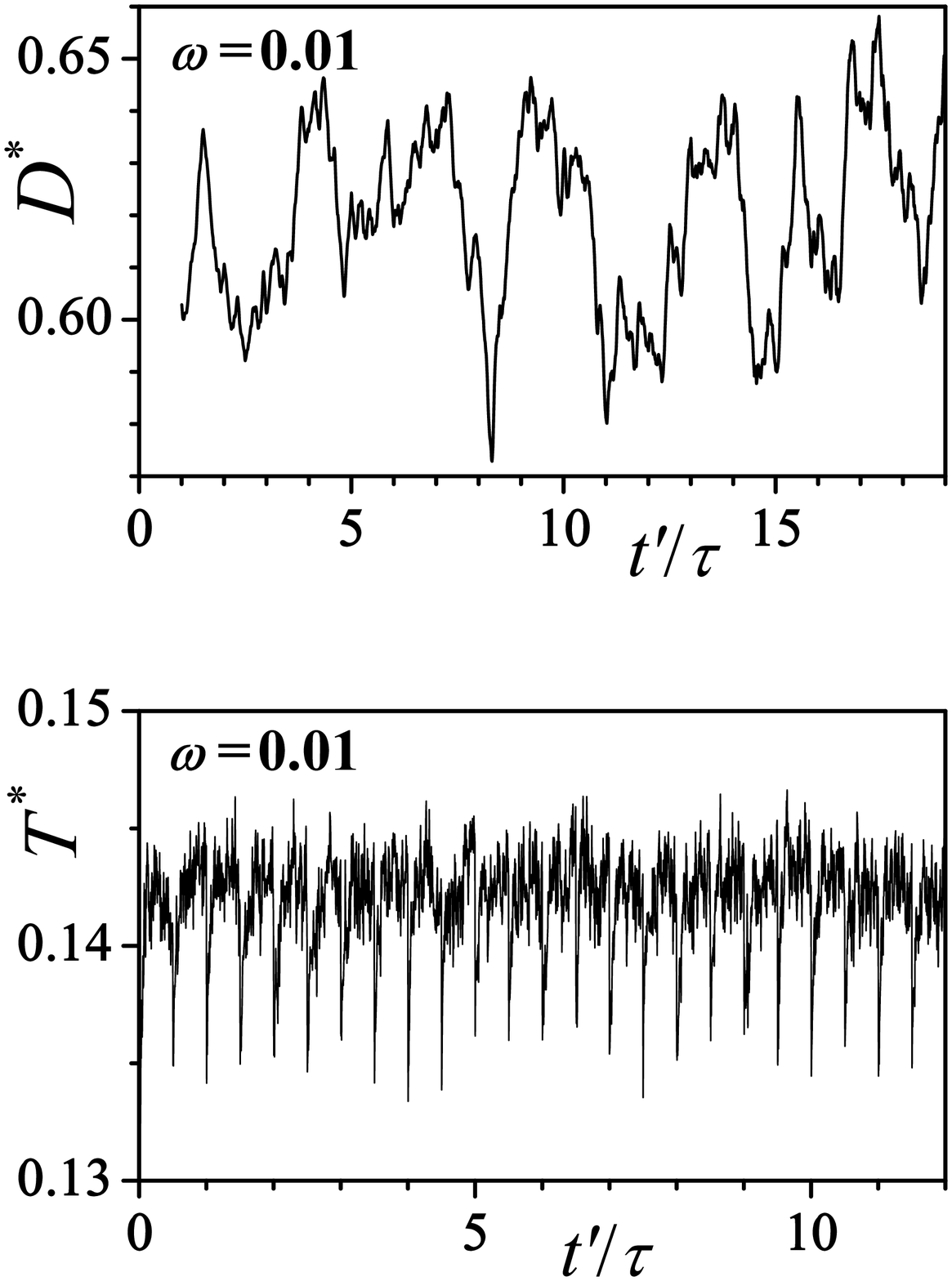}
       \caption{Evolution of the diffusivity and the kinetic temperature for
       high and small frequencies of the forcing, $\omega=2$ and $10^{-2}$.
       $F_e/F_0=0.06$. Note that time normalized to one \emph{forcing} period
       $\tau$ (rather than to $\mathcal{T}_0$ as in the previous $t$-dependence
       graphs) is shown along the abscissa axis.}
    \label{f:Tkin}
  \end{center}
\end{figure}

To analyze this we introduce kinetic temperature
$T^*=m\langle\Delta v^2\rangle/U_0$
(\emph{i.e.}, made dimensionless with the same multiplier $k_B/U_0$ as
$T^\dagger$), and study its evolution at different $\omega$'s:
Fig.~\ref{f:Tkin}, bottom. At low frequency, $T^*$ does not change
(only shows random noise) after one period $\tau=2\pi/\omega$ of the forcing.
At larger $\omega= 2$ regular oscillations in
$T^*$ are observed at all times, with amplitude of order 5\% of the average
$T^*$ value. That average is about 10\% higher than the average $T^*$ at
$\omega=10^{-2}$. $T^*$ averaged over the forcing oscillation period grows
for about $t^\prime_\textrm{st}\approx 7\tau$ until it reaches its stationary
value.

Together with $T^*$, $\sigma^2(t^\prime)$ also shows transient behavior
at $t^\prime < 7\tau$. We illustrate this in the top graphs in
Fig.~\ref{f:Tkin}, by plotting
\begin{equation}\label{Dstar}
  D^*(t^\prime)=\lbrack\sigma^2(t^\prime+\tau)-\sigma^2(t^\prime)\rbrack/(2\tau).
\end{equation}
In settled ``normal diffusion'' regime this quantity must coincide
(in average over the forcing period of duration $\tau$) with the
diffusivity, as in such a regime theoretically
$\sigma^2(t+n\tau)=\sigma^2(t)+n\sigma^2(\tau)$ for $n\in\mathbb{N}$.

These results in part mean that the rule of superdiffusion or dispersionless
stage (however these are modified in the high-frequency regime) terminating
and turning into normal diffusion
after one period of forcing no longer applies at high forcing
frequencies. This leads to the $D(\omega)$ derivation based on approximations
Eqs.~(\ref{sigma2fitI},\ref{sigma2fitIII}) progressively less accurate at
higher $\omega$'s, the power-law decay first slowing down, and then going
into a completely different regime at $\omega$ approaching $2\pi$.

At $\omega>2\pi$ the diffusivity tends to the diffusivity of particles in the
lattice periodic potential in the absence of extra driving $F_t(t)$.
Such an asymptote is shown by the horizontal dot-dashed line in
Fig.~\ref{f:Domega0095}.
\section{Dependence of diffusivity on temperature at fixed forcing
frequency}\label{s:DvsT}
Having studied how the temperature behavior of diffusion
changes at gradual forcing frequency increase, here we investigate in more
detail how the diffusivity changes with the temperature at fixed
frequency.
Besides theoretical importance, the results are of interest to
experimentalists. Oftentimes changing the frequency in the system is
problematic, while changing the temperatures is simpler; so the findings
in this section are more straightforward to test.
\begin{figure}[htbp!]
  \begin{center}
    $\!\!\!\!\!\!\!\!\!\!\!$\includegraphics[width=0.238\textwidth]{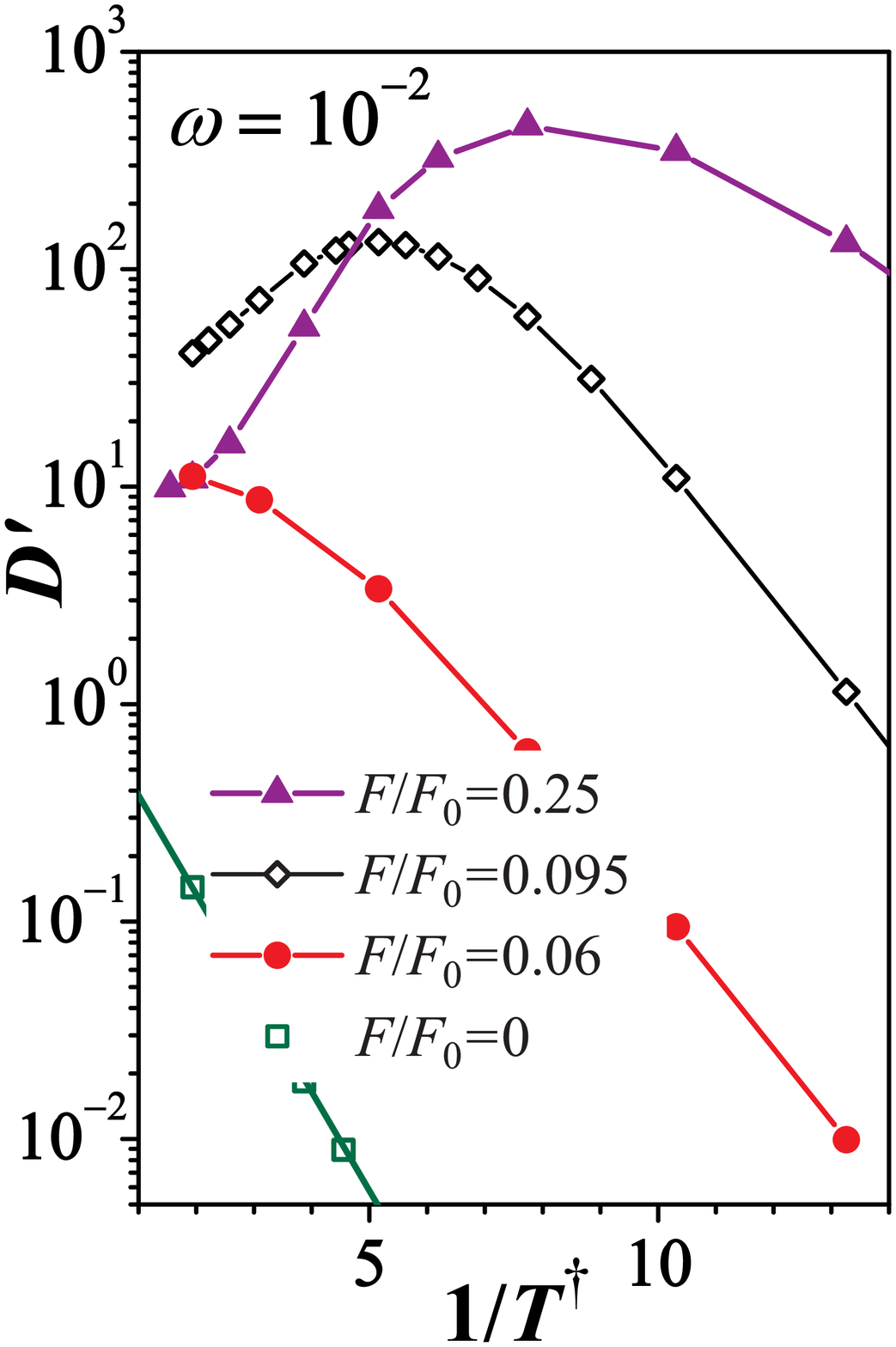}
    {\includegraphics[width=0.24\textwidth]{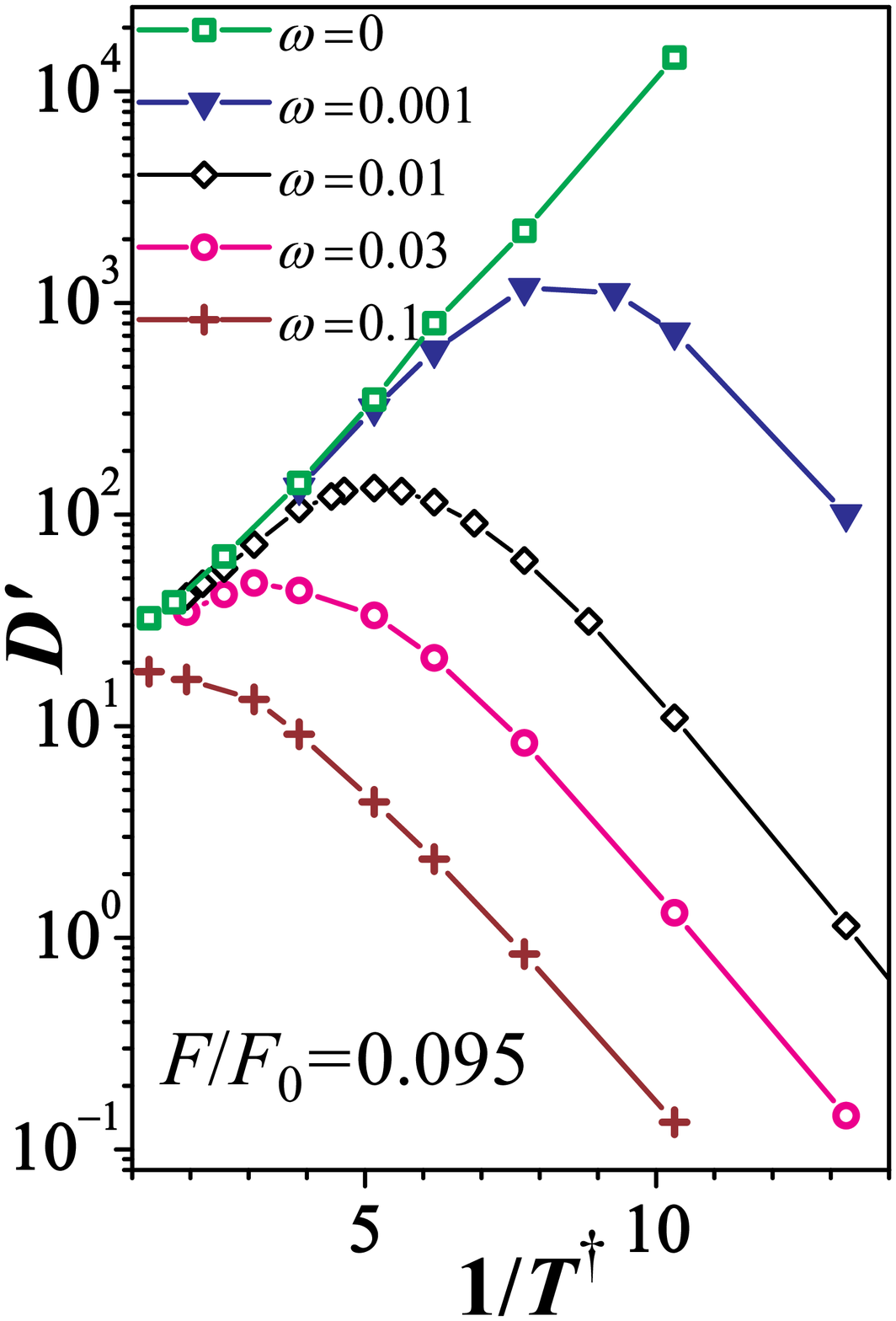}%
     $\!\!\!\!\!\!\!\!\!\!\!\!$}
       \caption{(Color online) Dimensionless diffusivity vs $1/T$ at \\
       (left): $\omega=10^{-2}$, force amplitudes $F_e/F_0=$ 0 (squares, and
       linear fit), 0.06 (circles), 0.095 (diamonds), 0.25 (triangles); \\
       (right): $F_e/F_0=0.095$, top to bottom: $\omega=$ 0 (squares),
        $10^{-3}$ (triangles), $10^{-2}$ (diamonds),
        $3\times 10^{-2}$ (circles), $10^{-1}$ (pluses).}
    \label{f:DTomega0001F}
  \end{center}
\end{figure}
\begin{figure*}[htbp!]
  \begin{center}
    $\!\!\!\!\!\!\!\!\!$\includegraphics[width=.343\textwidth]{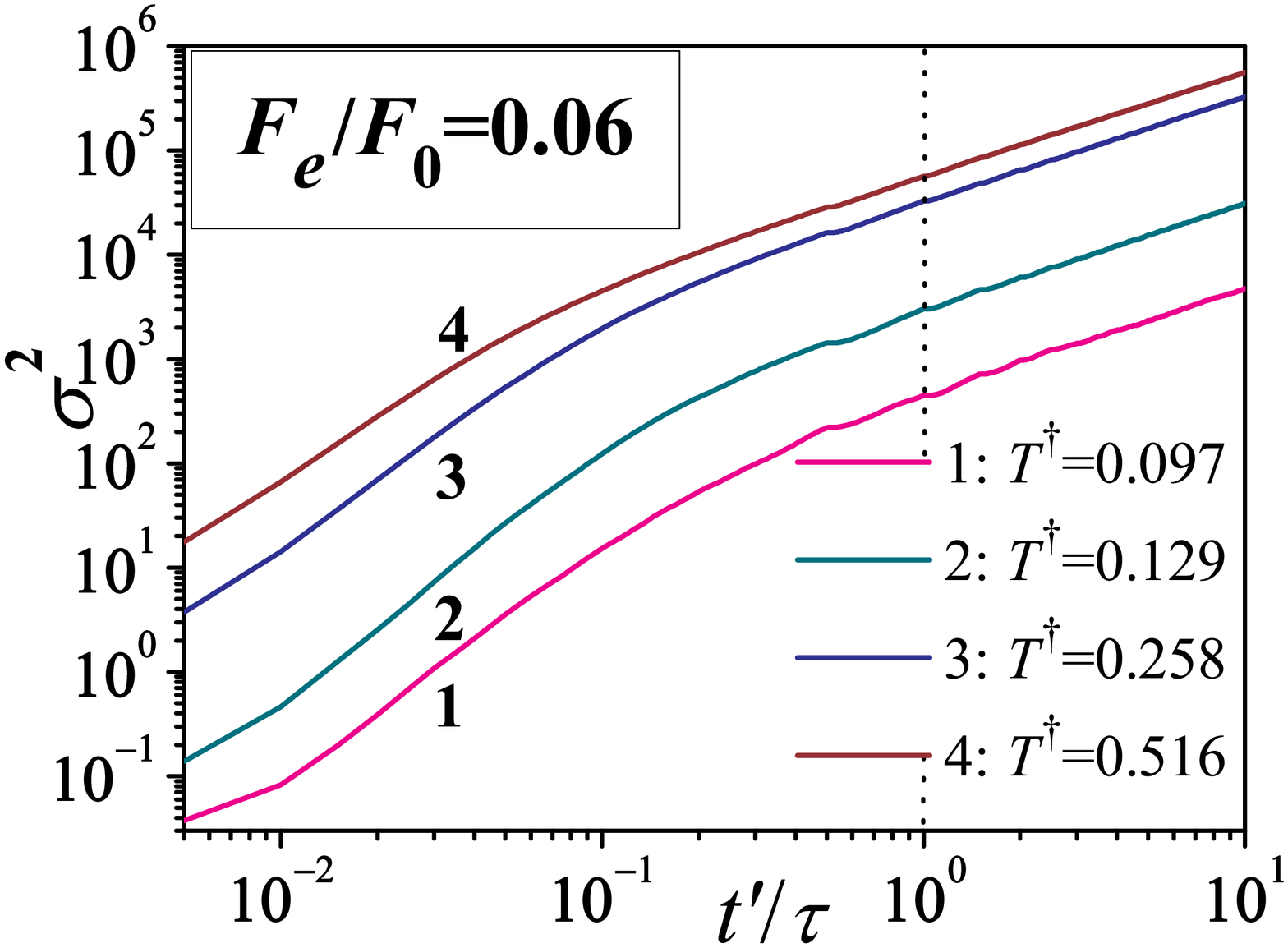}
    $\!\!\!\!\!$\includegraphics[width=.343\textwidth]{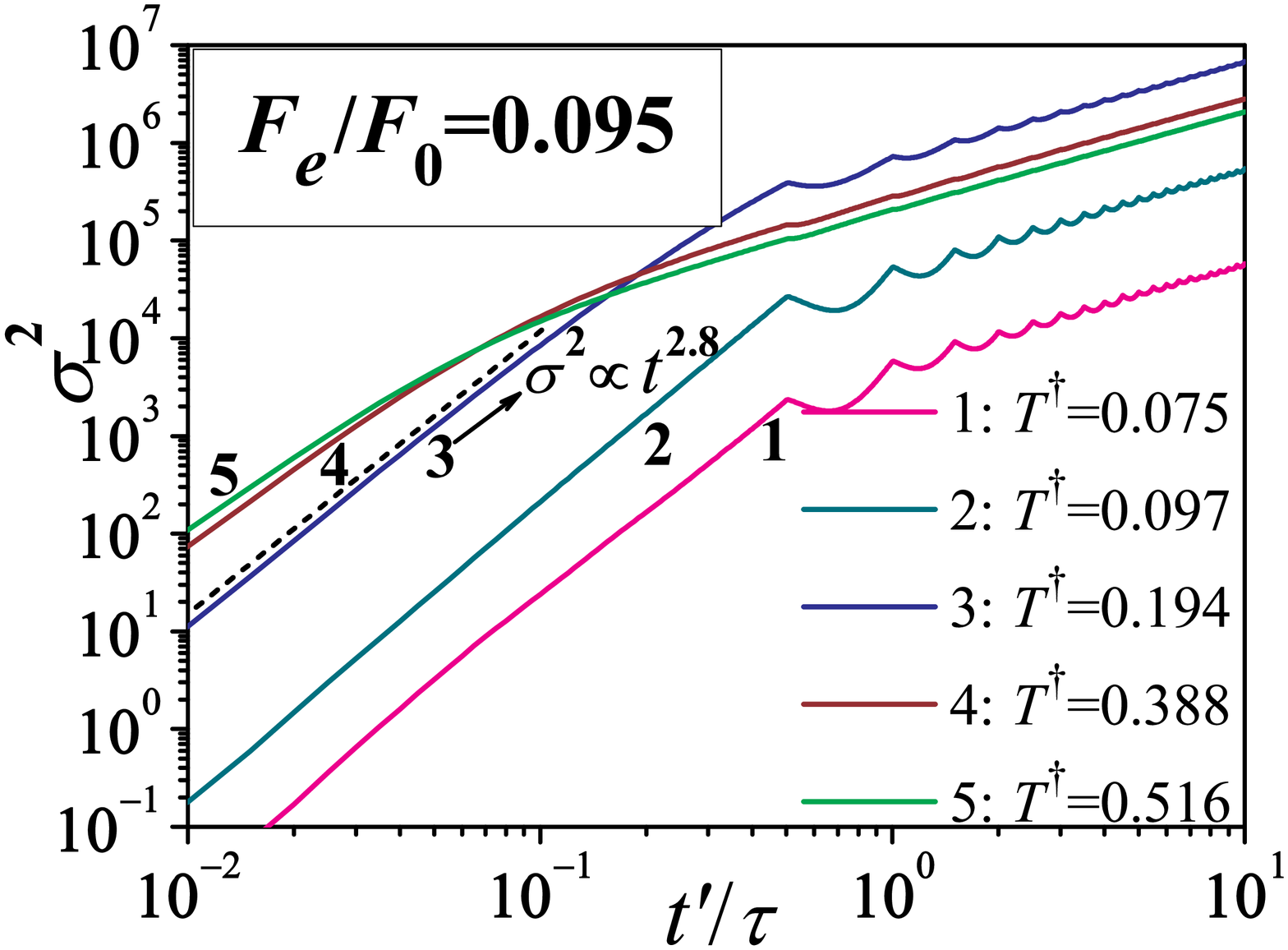}
    $\!\!\!\!\!$\includegraphics[width=.3465\textwidth]{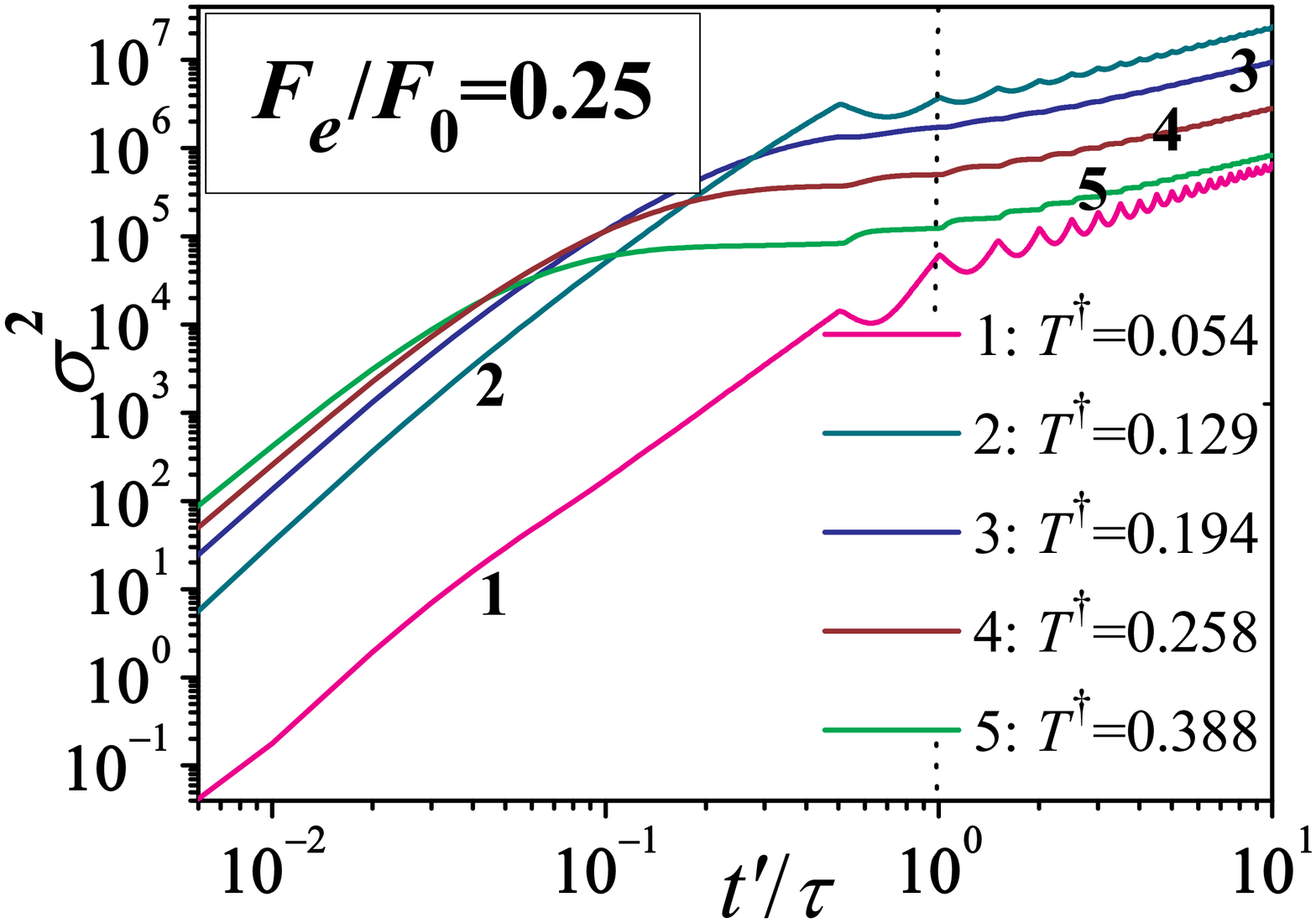}%
    $\!\!\!\!\!\!\!\!$
    \caption{(Color online) Dispersion of the particles as a function of time
    at different temperatures. Force amplitudes in regions I--III.
    $\omega=10^{-2}$.}
    \label{f:sigma3F}
  \end{center}
\end{figure*}

Left part of Fig.~\ref{f:DTomega0001F} shows how the diffusivity changes with
$T^\dagger$ at fixed $\omega = 10^{-2}$. For the forcing amplitude in region I
the diffusivity monotonically grows with the temperature, this (normal) behavior
agreeing with such at constant forcing \cite{MarchenkoFconst}. For $F_e$ in
regions II and III TAD is observed, as already noted in
Sec.~\ref{s:RegionII}--\ref{s:RegionIII}. TAD is confined to limited
(from below) temperature intervals, contrary to the TAD behavior observed in
\cite{MarchenkoFconst} for constant forcing. In \cite{MarchenkoFconst} the
diffusivity was numerically checked growing (only in region II at
$\omega=0$)
at temperature decreasing down to $T^\dagger=0.097$. Such a behavior was
understood theoretically, dependence $D\propto \exp[\mathcal{E}/(k_BT)]$ was
predicted at low temperatures going all the way down to 0, and agreed well with
simulation results.

In the current setup with periodic forcing, similar temperature dependence is
observed too,
\begin{equation}\label{DmidT}
  D\propto \exp(\epsilon_1/T^\dagger) ,\; \epsilon_1>0,
\end{equation}
but only above certain temperature. The growth slows down as the
temperature keeps decreasing, $D$ reaches maximum $\Dmax(\Tmax)$ at certain
$\Tmax$, and starts decreasing with further decrease in $T$. The behavior at
lower temperatures is well approximated by
\begin{equation}\label{DlowT}
D\propto \exp(-\epsilon_2/T^\dagger) ,\; \epsilon_2>0.
\end{equation}
In the absence of forcing we observe $\epsilon_2=1$ (corresponding dimensionful
quantity $\mathcal{E}_2=U_0$) --- classical Arrhenius temperature behavior of
the diffusivity.
$\epsilon_2$ decreases with $F_e$, whereas $\epsilon_1$ increases.
$\Tmax$ and $\Dmax$ increase with $F_e$ as well.

Temperature behavior of the diffusivity is related to such of the
superdiffusion stage duration $\tau_\textrm{spd}$ and $\tau_1$. These times
may be inferred from simulations of the particle dispersion evolution with
time,
Fig.~\ref{f:sigma3F}. For the forcing amplitude in region I, $\tau_1$ (the
time at
which superdiffusion starts, about the potential well escape time) increases
with the temperature decreasing. $\tau_\textrm{spd}=\tau_2-\tau_1$ at the same
time decreases. This is seen to result in the diffusivity growing with the
temperature, agreeing with the results shown in the
$\omega=10^{-2}$ part of Fig.~\ref{f:DTomega0001F}.

In region II $\tau_1$ increases with the temperature decreasing as well. %
$\tau_\textrm{spd}$ however also increases with the temperature decreasing
(middle graph in Fig.~\ref{f:sigma3F}, curves 3--5) --- contrary to the
behavior for $F_e$ in region I. TAD is realized at these temperatures. At
temperature
$\Tmax(F,\omega)$ ($T^\dagger\approx 0.194$ in this graph) $\tau_2$ becomes
equal to $\tau=2\pi/\omega$, the period of forcing. At this temperature the
maximal diffusivity is achieved, corresponding to the longest actual
superdiffusion stage at the given forcing frequency.
TAD is only realized at temperatures above $\Tmax(F_e,\omega)$: even though
$\tau_\textrm{spd}$ (superdiffusion duration at constant forcing) still grows
at further temperature decrease together with $\tau_2$,
in periodic forcing setup superdiffusion
effectively ends earlier, at $t^\prime\approx\tau$. It is this superdiffusion
interruption at $t^\prime\approx\tau$ that makes $D(T)$ behavior at lower
temperatures to diverge in the periodic forcing setup from that at constant
forcing. In the constant forcing problem, based on the same logic, TAD must be
realized all the way down to $T=0$; in agreement with conclusions in
\cite{MarchenkoFconst}.


It is clear from this argument that there is no uniform convergence with
$\omega$ of $D(T)|_\omega$ curves to $D(T)|_{F=const}$.
At $\omega$ getting smaller $D(T)|_\omega$ gets close to such a $D(T)$ curve
for constant forcing on longer $T$-interval, down to progressively lower
temperatures. However at any $\omega$ there exists $T_\textrm{TAD}(\omega)$
such that normal temperature dependence of the diffusivity is restored at
$T<T_\textrm{TAD}(\omega)$.
This non-uniform convergence is shown in the right plot in
Fig.~\ref{f:DTomega0001F}.

The same behavior is observed in region III. The maximal diffusivity is
achieved at $\Tmax(F_e,\omega)$ corresponding to the longest superdiffusion
stage at the given $\omega$. The lowest temperature at which TAD is realized is
the one at which $\tau_2$ reaches the period of external force $\tau$ (more
precisely, the time of superdiffusion interruption at a given $\omega$).
At this temperature $\partial D/\partial T|_{\{F_e,\omega\}=const}$ crosses
zero value. The interval of temperatures, in which TAD is observed,
is wider than at forcing in region II, due to the presence of dispersionless
phase (curve 4 in the right plot in
Fig.~\ref{f:sigma3F}). Experimentally it is simpler to observe TAD at larger
$F_e$, at which within the period of $F_t(t)$ the superdiffusion stage is used
to larger extent; and the temperature interval of TAD is thus wider.

\begin{figure*}[htbp!]
  \begin{center}
    $\!\!\!\!\!$\includegraphics[width=.336889\textwidth]{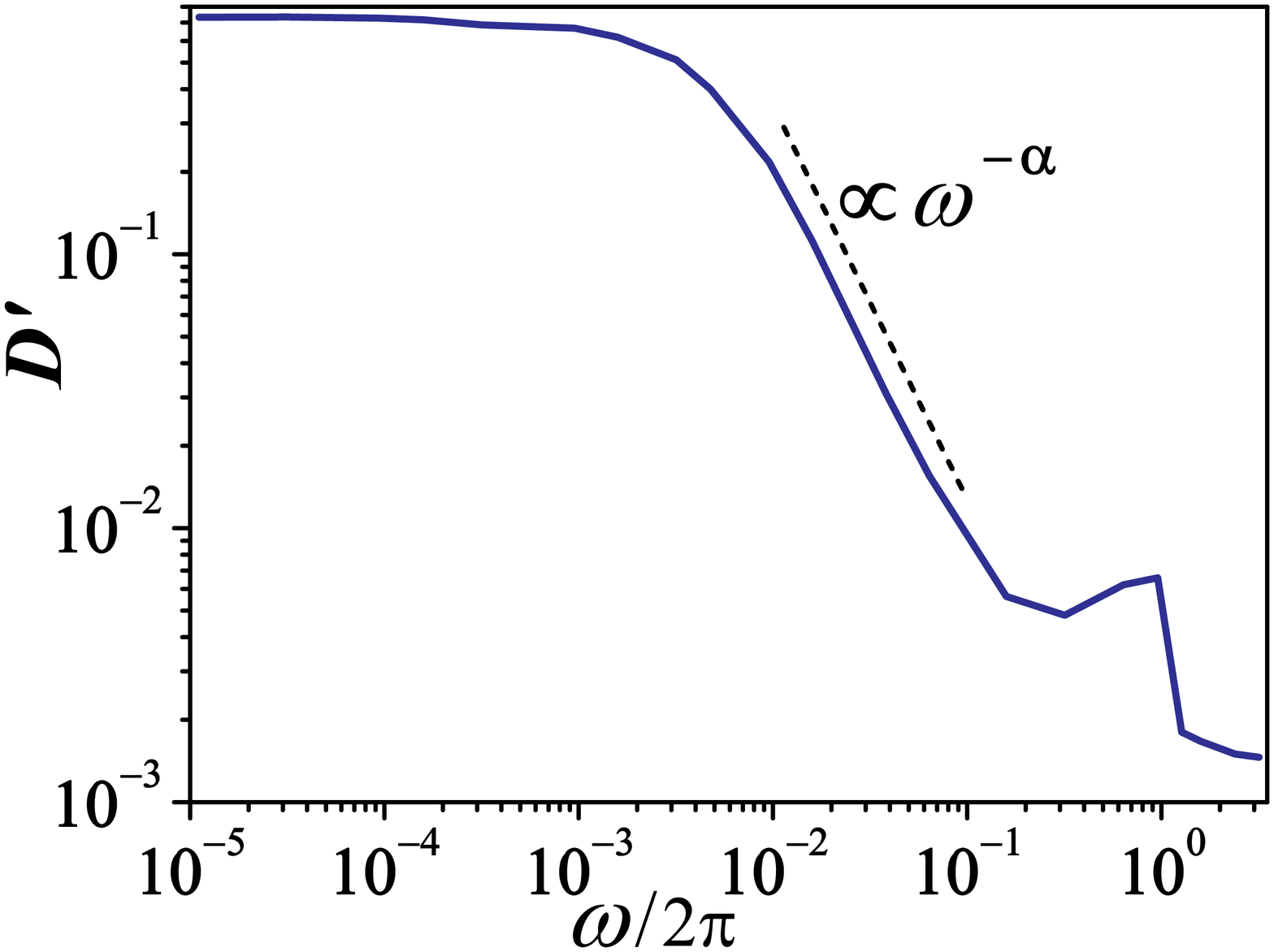}
    $\!$\includegraphics[width=.3345\textwidth]{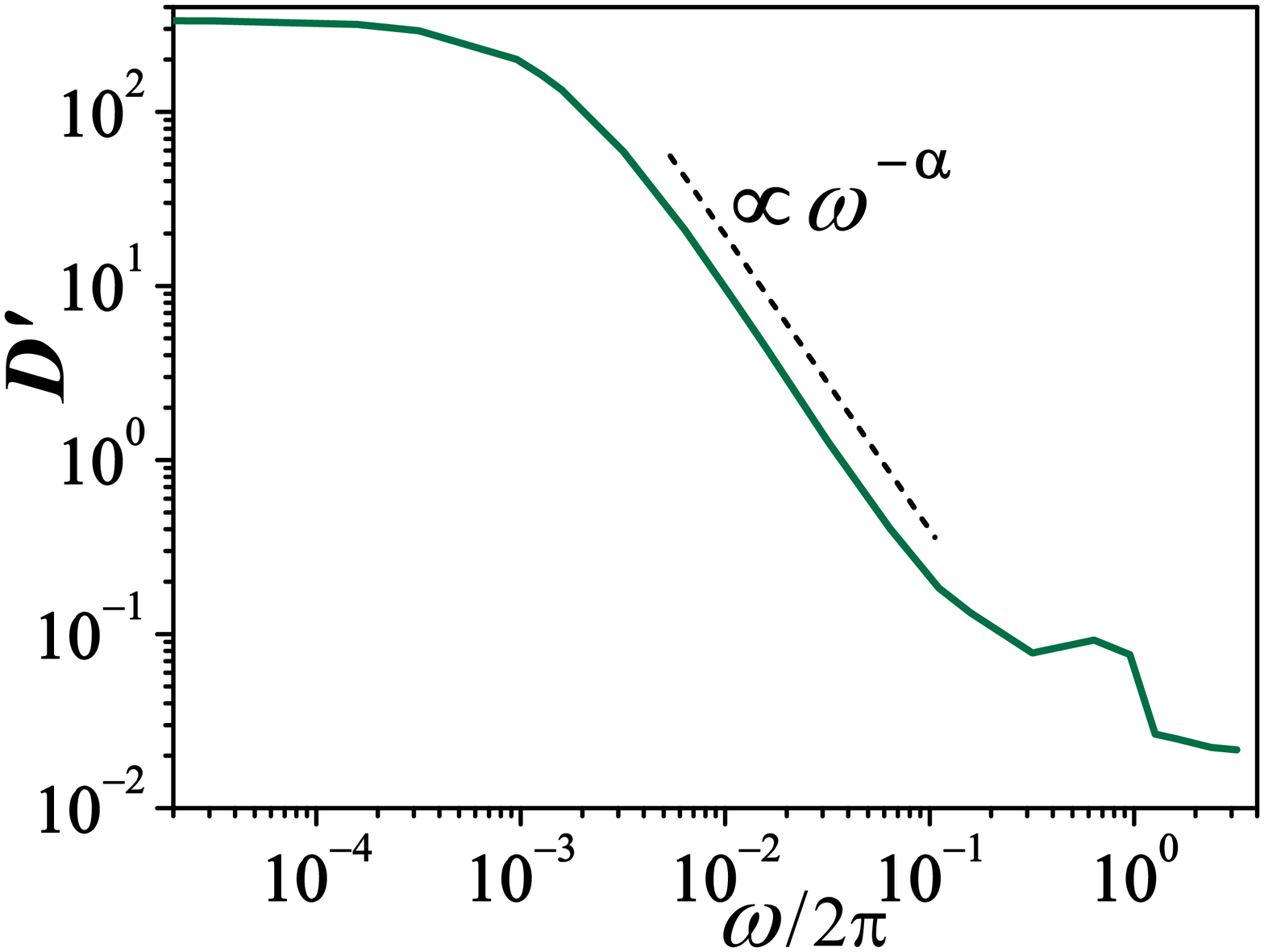}
    $\!$\includegraphics[width=.330073\textwidth]{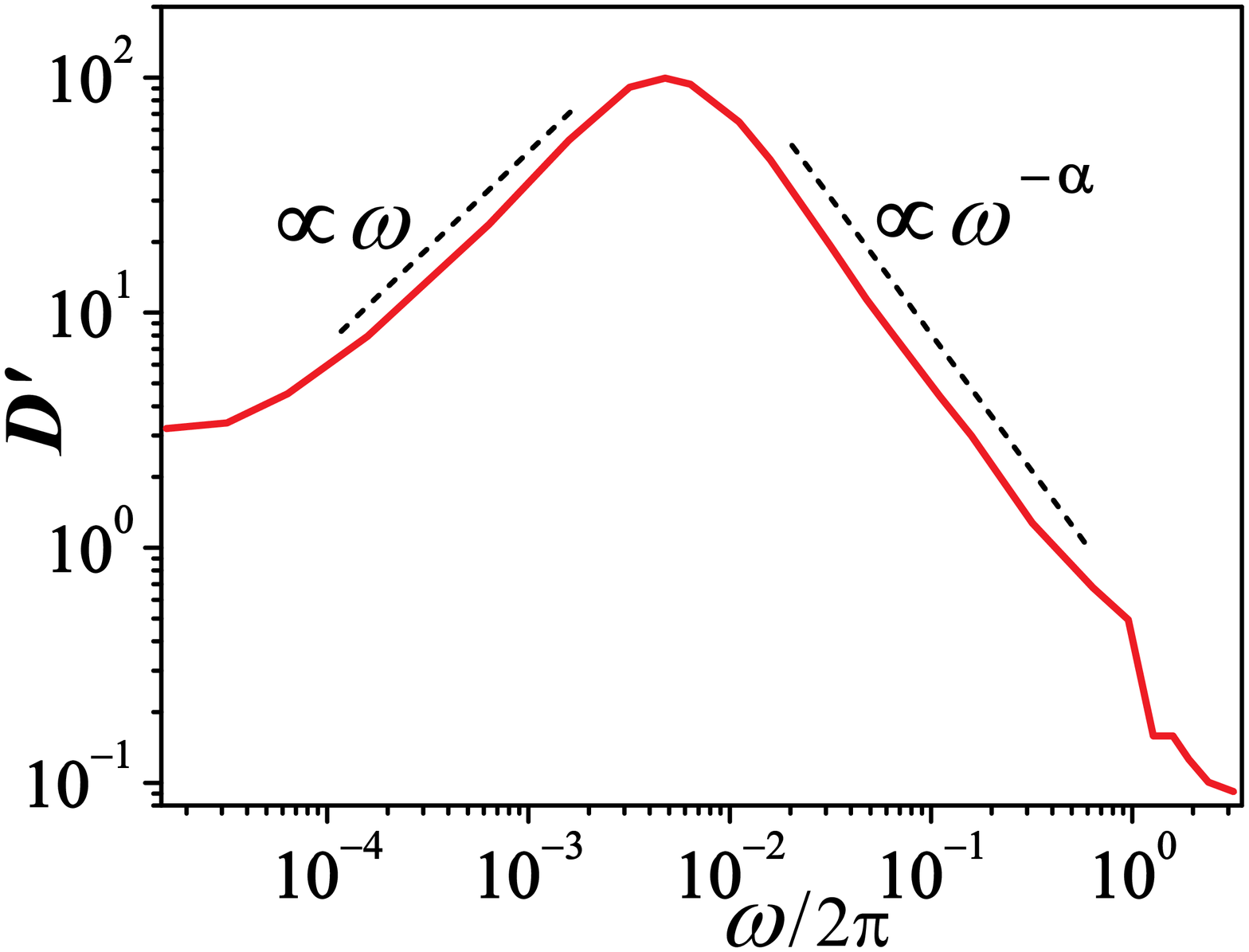}%
    $\!\!\!$\\
    $\!\!\!$\includegraphics[width=.3334\textwidth]{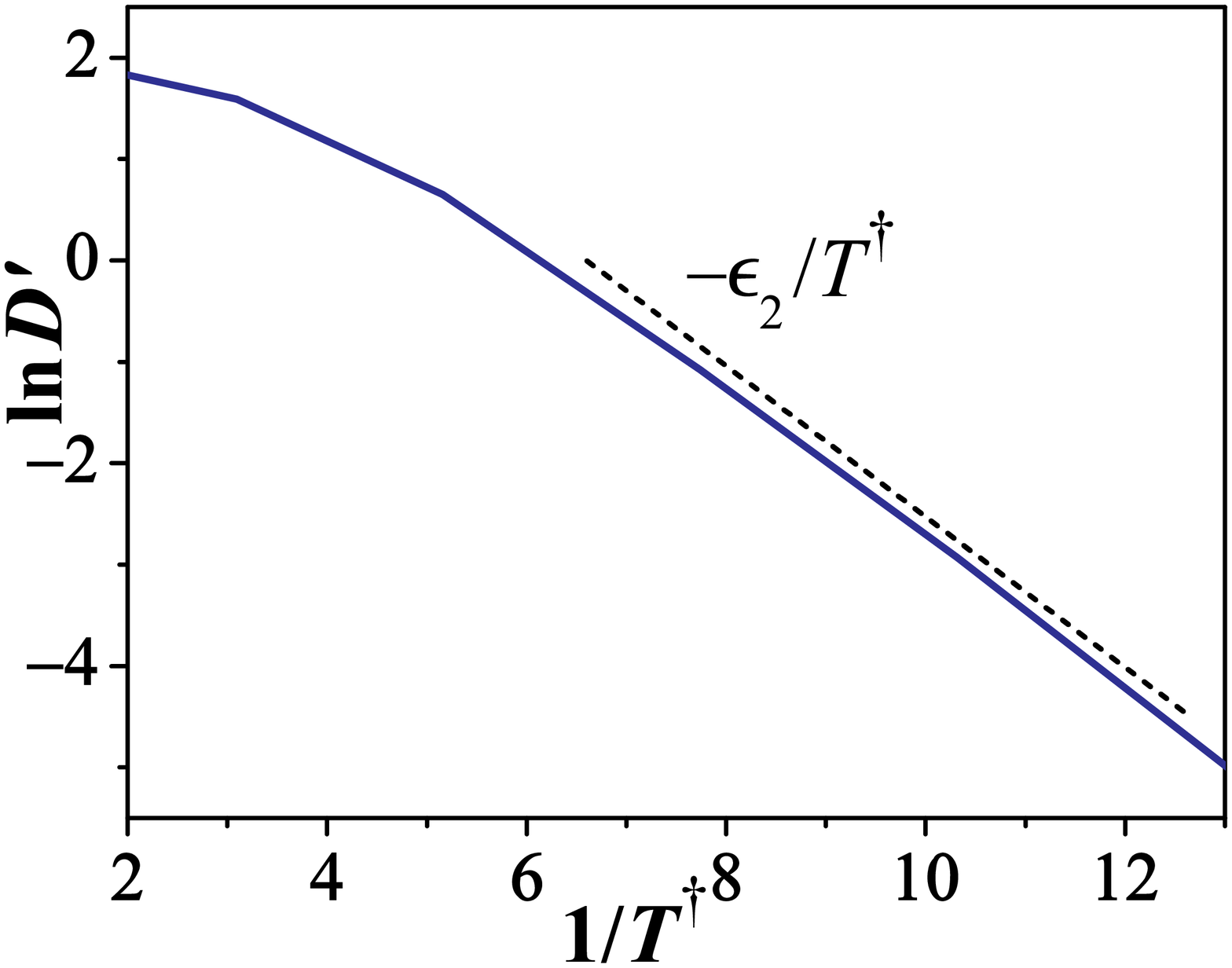}
    $\!\!$\includegraphics[width=.338402\textwidth]{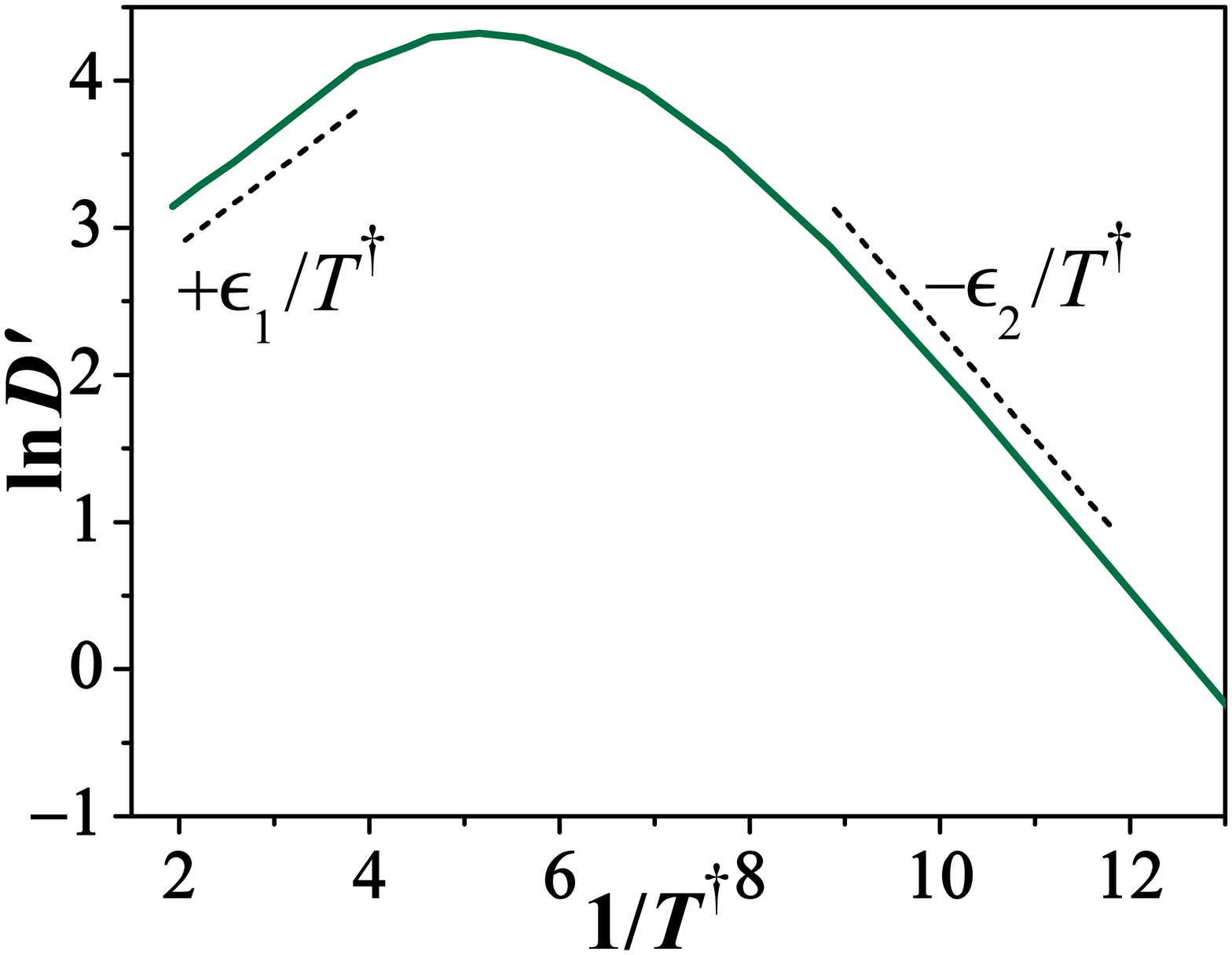}
    $\!\!$\includegraphics[width=.338402\textwidth]{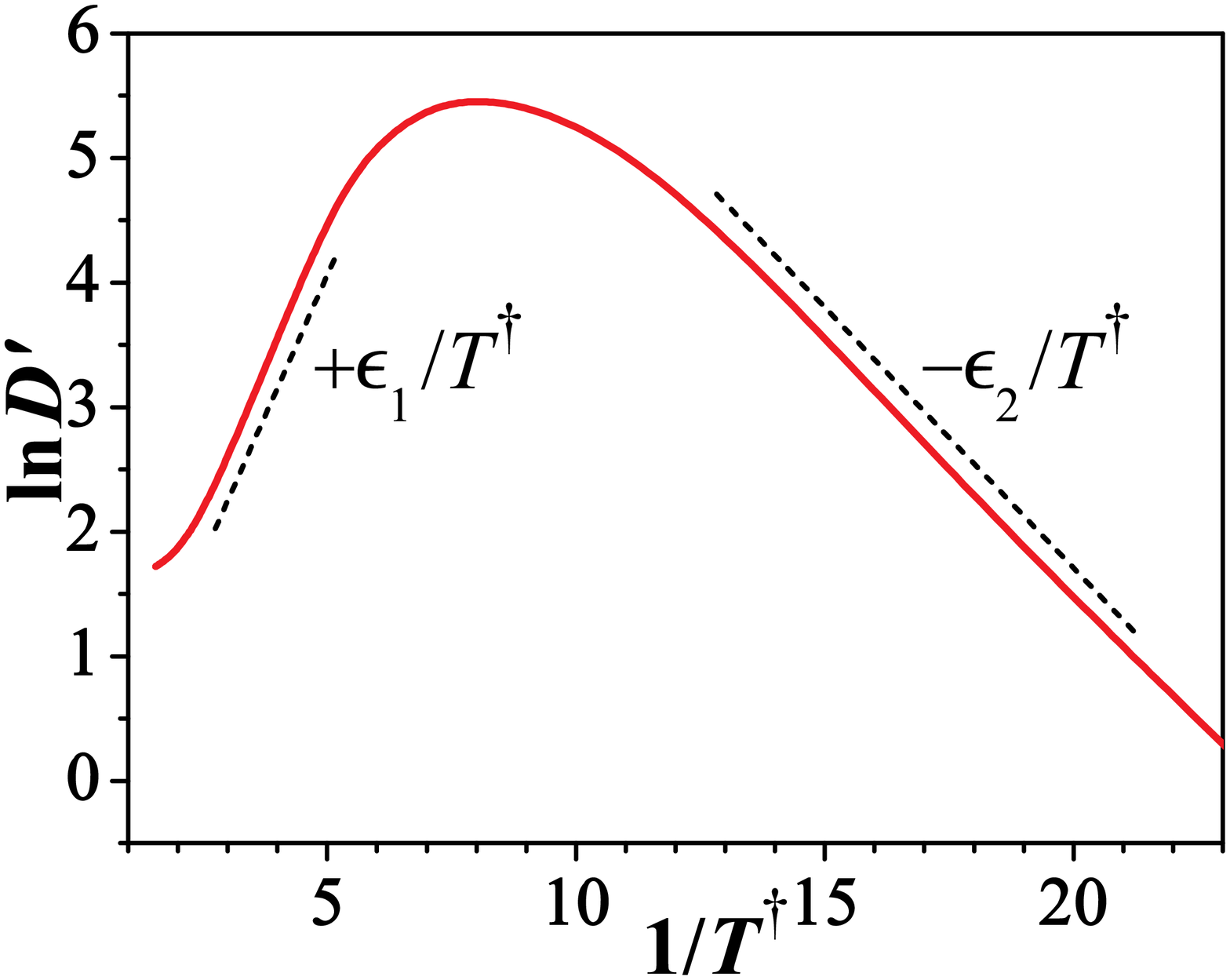}%
    $\!\!\!$
    \caption{(Color online) Features of the dependence of the diffusivity on
    the forcing frequency at fixed temperature (top) and on the inverse
    temperature at fixed frequency (bottom), for $F_e$ in regions I--III.}
    \label{f:DTomega}
  \end{center}
\end{figure*}
Presented study of the diffusion under external forcing can be summarized by
the following schematics, Fig.~\ref{f:DTomega}. Force amplitudes fall into
three regions, I--III. In region I the diffusivity grows with the temperature
increasing. In regions II--III the dependence $D(T)$ is nonmonotonic, it
shows a maximum at certain $\Tmax$. Below $\Tmax$ the temperature dependence
of the diffusivity is normal, $\partial D/\partial T>0$. Temperature interval
of TAD exists above $\Tmax$, $\partial D/\partial T<0$.

In the forcing frequency dependence of the diffusivity, on the other hand, it
is regions I and II that look alike: at the frequency decreasing from
a fraction of the potential eigenfrequency the diffusivity grows monotonically,
getting to a flat region at $\omega<\omega_2$, in which it virtually reaches
its limiting $D(\omega=0)$ value. In region III the $D(\omega)$ curve passes
through a maximum, near $\omega=\omega_2\equiv 2\pi/\tau_2$, corresponding to
the end time of superdiffusion stage $\tau_2$ at constant forcing. At
frequencies getting smaller, the forcing period contains progressively
longer interval of ``would-be'' (in the constant forcing problem)
dispersionless and (at yet smaller frequencies) normal-diffusion time-interval,
resulting in $D$ decreasing at such further $\omega$ decrease.
\begin{figure*}[htbp!]
  \begin{center}
    $\!\!\!\!\!\!\!$\includegraphics[width=.3368\textwidth]{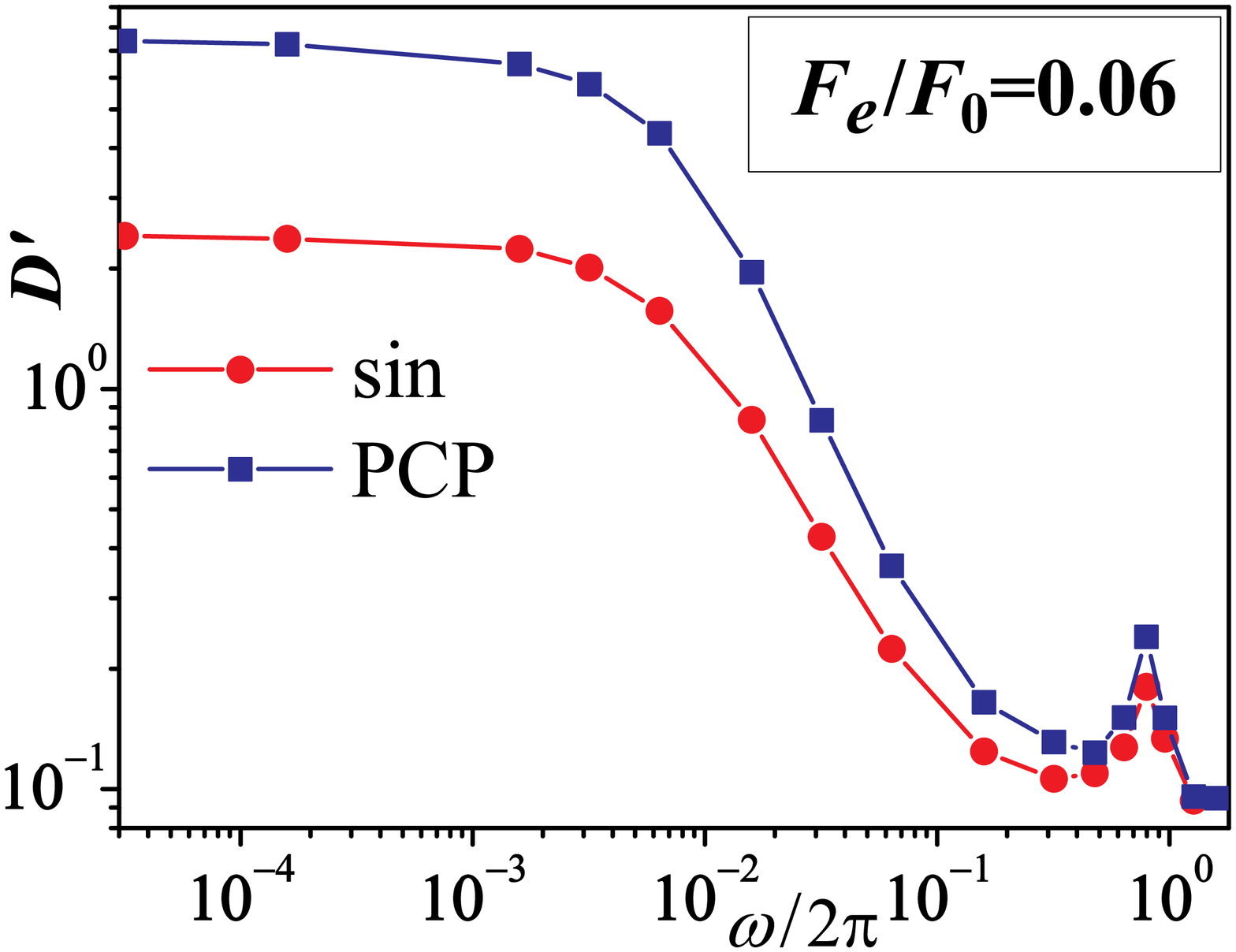}
    $\!\!\!$\includegraphics[width=.340999\textwidth]{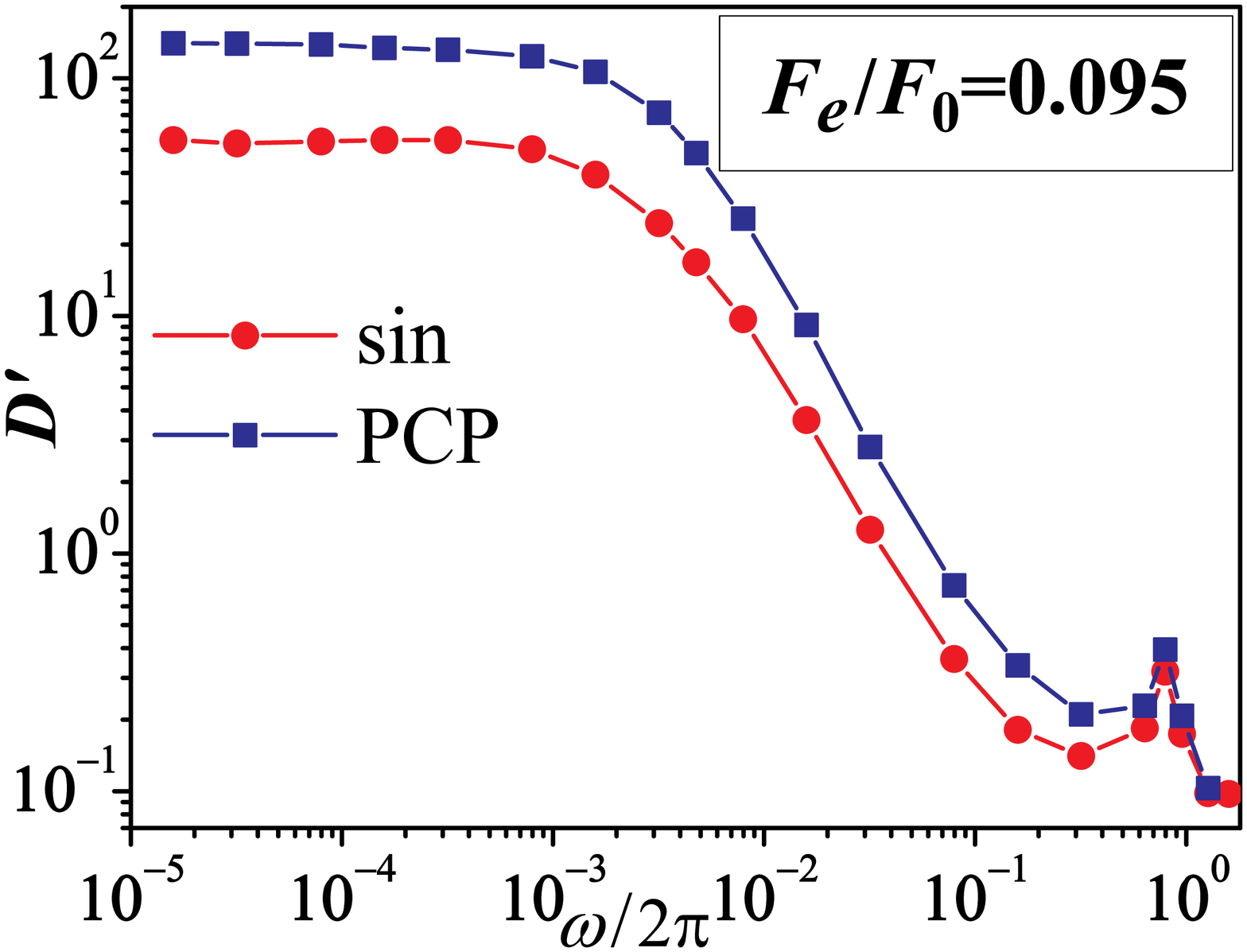}
    $\!\!\!$\includegraphics[width=.337291\textwidth]{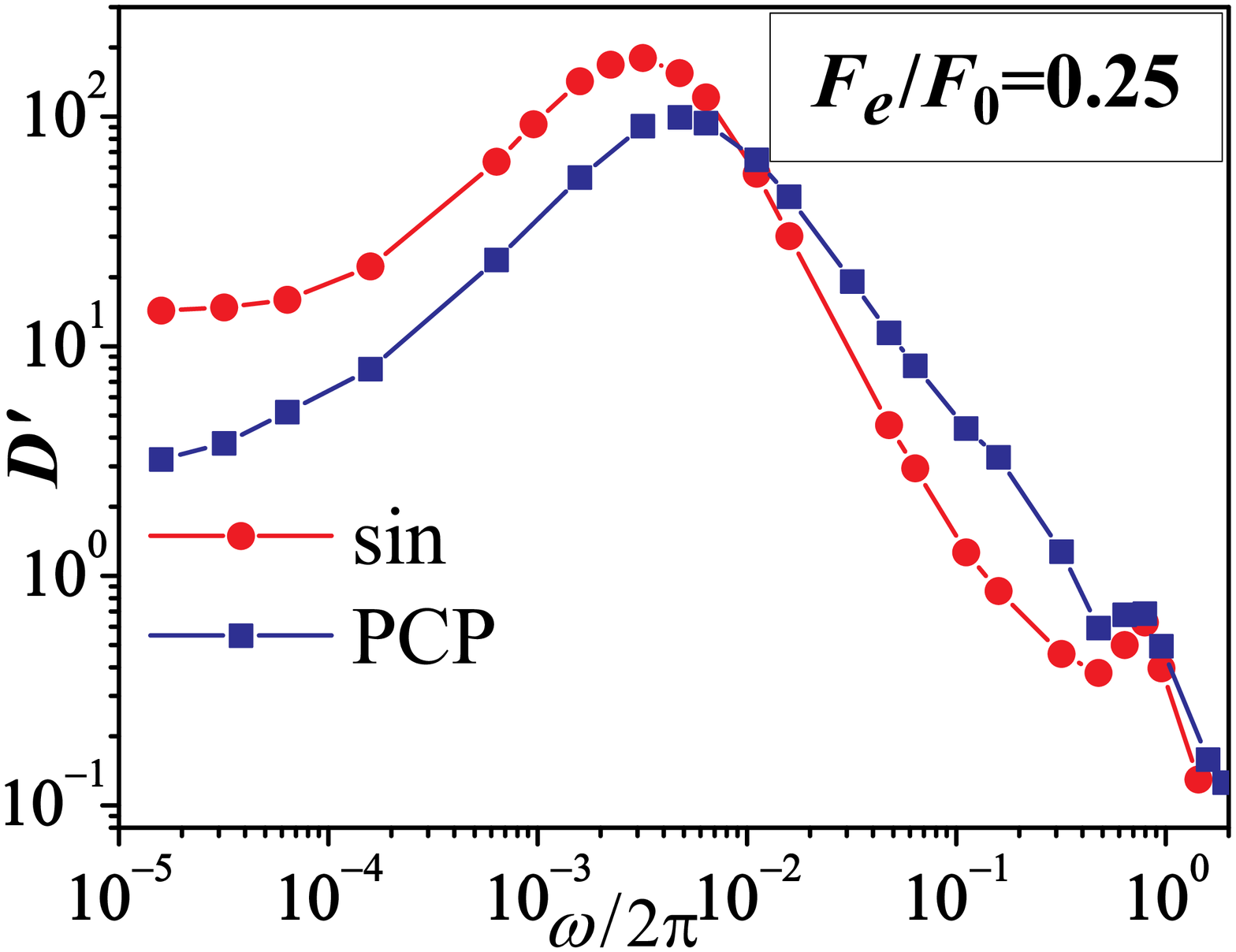}%
    $\!\!\!\!\!\!\!$
    \caption{(Color online) $D^\prime(\omega)$ for force amplitude in regions I--III. $T^\dagger =0.258$. Sinusoidal in time (circles) and PCP (squares) forcing. }
    \label{f:DFsin}
  \end{center}
\end{figure*}

\section{Sinusoidal in time external forcing}\label{s:Fsin}
Here we show the frequency dependence of the diffusivity for sinusoidal in time
external forcing.
We observe the same features in $D(\omega)$ as we saw for PCP forcing.
Main conclusions of the sections above thus appear not to depend critically
on specific form of time-dependence of the (smooth enough, symmetric) forcing.

Figure~\ref{f:DFsin} presents a comparison between diffusivities at sinusoidal
in time and PCP forcing. At small frequencies the diffusivity stays close to
its value at $\omega=0$. In regions I and II of the external
forcing amplitude $D(\omega)$ starts decaying according to a power law at
$\omega>2\pi/\tau_2$, $\tau_2$ being the superdiffusion stage ending time at
constant external forcing with the same amplitude $F_e$. In region III
$D(\omega)$ grows linearly with
$\omega$ at intermediate $\omega$'s, reaches a maximum, then shows a power-law
decay at larger frequencies. For all force amplitudes the power law decay
switches to more complex behavior at $\omega$ reaching about $1/10^\textrm{th}$
of the potential eigenfrequency.

Presence of periodic forcing of any form of time-dependence leads to
significant enhancement in diffusivity. This enhancement is somewhat smaller
for sinusoidal forcing in regions I, II (can be understood, as at the same
$F_e$ the average over period forcing $\langle |F_t|\rangle_t$ is smaller
for sinusoidal than for PCP forcing). In region III the maximum in
diffusivity is achieved at smaller $\omega_\textrm{max}$, and the value of
$\Dmax$ is larger than such a maximal diffusivity at corresponding
$\omega_\textrm{max,PCP}$ under PCP driving with the same amplitude $F_e$.
This can be understood from the same argument, as in region III the
diffusivity \emph{decreases} with the forcing increasing, all other
parameters kept constant.%
\section{Conclusions}\label{s:concl}

We investigated enhancement of diffusion in 1D space-periodic underdamped
systems by external time-periodic fields. We showed that the diffusion can
be enhanced by orders of magnitude at certain choice of temperatures $T$,
forcing amplitudes $F_e$ and frequencies $\omega$. Three regions of $F_e$
exist, I--III, in which dependences of diffusivity $D$ on $T$ and $\omega$
differ qualitatively.

At all $F_e$ there is an interval of small frequencies, in which $D$ depends
only weakly on the frequency, approaching the limiting value $D(\omega=0)$.
This value coincides with $D$ at constant bias force with the same value $F_e$
in the case of piecewise constant in time periodic (PCP) forcing, the form
$F_t(t)$ of forcing we focused on in this study. At larger frequencies, from
$\omega_2=2\pi/\tau_2$ corresponding to the superdiffusion stage ending time
$\tau_2$, to about $2\pi/10$ (dimensionful frequency
$\Omega\approx 1/10^\mathrm{th}$ of the potential eigenfrequency
$\Omega_0$) $D(\Omega)$ decays with $\Omega$ according to a power-law.
The exponent of this decay is related to the superdiffusion stage exponent
$\alpha$. At yet larger frequencies non-equilibrium effects slow down the
power-law decay of $D(\Omega)$, and, after a local maximum in $D(\Omega)$ near
$\Omega\approx 0.8\Omega_0$, $D$ tends to its asymptotic value coinciding with
the diffusivity in the absence of external forcing.

At small forcing amplitudes, in regions I and II, $D(\Omega)$ decreases
monotonically with frequency $\Omega$ from its value at constant forcing, all
the way till the non-equilibrium effects become critical,
at $\Omega\approx 0.3\Omega_0$. At stronger forcing, region III, $D(\omega)$
is nonmonotonic, a maximum is reached at
$\omega_\textrm{max}\approx\omega_2$. $D(\omega)$ grows monotonically on
$\omega\in(0; \omega_\textrm{max})$, this growth is approximately linear on
certain interval to the left of $\omega_2$.

We studied the temperature dependence of the diffusivity at fixed $\omega$.
Contrary to the constant forcing problem \cite{MarchenkoFconst} the diffusivity
increases with the temperature for all $F_e$ when the temperature is below
certain frequency-dependent threshold value. Limited temperature intervals
exist for the force amplitudes in regions II and III, in which
the diffusivity decreases with the temperature
(``temperature-abnormal diffusivity'', TAD).

The physical reason behind the strong diffusivity enhancement is emergence
of two populations of particles under the action of external force, locked and
running ones. At optimal $F_e$ and $T$ the number of particles in the two 
populations is comparable,  long flights of the running particles
relative to the locked population take place with significant probability,
resulting in giant enhancement of diffusion.

For comparison we studied diffusivity under sinusoidal in time driving.
We saw that the same features in $D(\omega)$ are observed in the three regions
of force amplitudes as in PCP driving case. We thus conjecture that the
qualitative features of $D(\omega, T, F_e)$ behavior are insensitive to
specific functional dependence of the (symmetric, smooth enough in $t$)
external force on time.

The effects investigated allow for simple experimental verification. Diffusion
of particles on solid body surface is one natural arena for this. Such systems
are characterized by low dissipation \cite{Bruch07RMP,*Krim12AdPhy}.
Another field to study
TAD in is in propagation of magnetic particles on non-magnetic substrate,
acted upon with electromagnetic fields. Abnormal diffusion enhancement could
be manifested in, \emph{e.g.}, enhanced growth of islands of the new phase at
decreasing temperatures.

Once verified experimentally, the effect of abnormal diffusion enhancement can
find applications in a number of new technologies: in sorting of particles,
manufacturing surface structures with required properties, controlling
penetration of particles through biological and artificial membranes,
in memristors, devices with charge density waves, etc.

\begin{acknowledgments}
We are grateful to Alexei Chechkin for useful discussions.
\end{acknowledgments}

\bibliographystyle{apsrev4-1}
%

\end{document}